\documentclass[11pt]{article}
\pdfoutput=1
\usepackage[utf8]{inputenc}
\usepackage[T1]{fontenc}

\usepackage{formatting}
\usepackage{shortcuts}

\begin{document} 
\begin{titlepage}
\begin{center}
\phantom{ }
\vspace{3cm}

{\bf \Large{Quantum chaos, integrability, and late times in the Krylov basis}}
\vskip 0.5cm
Vijay Balasubramanian${}^{1,2,3}$, Javier M. Magan${}^{4}$, Qingyue Wu${}^{1}$
\vskip 0.05in

\small{${}^{1}$ \textit{David Rittenhouse Laboratory, University of Pennsylvania}}
\vskip -.4cm
\small{\textit{ 209 S.33rd Street, Philadelphia, PA 19104, USA}}

\vskip -.10cm
\small{  ${}^{2}$ \textit{Santa Fe Institute 
}}
\vskip -.4cm
\small{\textit{1399 Hyde Park Road, 
Santa Fe, NM 87501, USA}}

\vskip -.10cm
\small{  ${}^{3}$ \textit{Theoretische Natuurkunde, Vrije Universiteit Brussel}}
\vskip -.4cm
\small{\textit{Pleinlaan 2,  B-1050, Brussels, Belgium}}

\vskip -.10cm
\small{  ${}^{4}$ \textit{Instituto Balseiro, Centro At\'omico Bariloche}}
\vskip -.4cm
\small{\textit{ 8400-S.C. de Bariloche, R\'io Negro, Argentina}}

\begin{abstract}
Quantum chaotic systems are conjectured to display a spectrum whose fine-grained features (gaps and correlations) are well described by Random Matrix Theory (RMT). We propose and develop a complementary version of this conjecture: quantum chaotic systems display a Lanczos spectrum whose local means and covariances are well described by RMT. To support this proposal, we first demonstrate its validity in examples of chaotic and integrable systems. We then show that for Haar-random initial states in RMTs the mean and covariance of the Lanczos spectrum suffices to produce the full long time behavior of  general survival probabilities including the spectral form factor, as well as the spread complexity. In addition, for initial states with continuous overlap with energy eigenstates, we analytically find the long time averages of the probabilities of Krylov basis elements in terms of the mean Lanczos spectrum. This analysis suggests a notion of eigenstate complexity, the statistics of which differentiate integrable systems and  classes of quantum chaos. Finally, we clarify the relation between spread complexity and the universality classes of RMT by exploring various values of the Dyson index and Poisson distributed spectra.

\end{abstract}
\end{center}

\small{\vspace{3.5 cm}\noindent ${}^{\dagger}$vijay@physics.upenn.edu\\
${}^{\ddagger}$javier.magan@cab.cnea.gov.ar \\
${}^{*}$ aqwalnut@sas.upenn.edu
}

\end{titlepage}

\setcounter{tocdepth}{2}

{\parskip = .4\baselineskip \tableofcontents}
\newpage

\section{Introduction}

Various aspects of chaotic systems can be approximately described by Random Matrix Theory (RMT). In particular, in many examples the distribution of gaps between nearby eigenvalues, the spectral form factor and  other spectral properties match those of random matrices \cite{Meh2004,Guhr:1997ve,bookhaake,akemann2011oxford}. This is generally not true for integrable and localized systems. Recently, there has been interest in studying quantum chaos by tridiagonalizing the Hamiltonian/Liouvillian, as opposed to directly studying the spectrum via diagonalization. Tridiagonalization of the Hamiltonian/Liouvillian with a fixed initial vector expresses it in the  corresponding Krylov basis, to be described below. The entries of the the resulting matrix are called the Lanczos coefficients, sometimes dubbed the Lanczos spectrum.

Part of the interest in this approach is that it has a closer bearing on the dynamics of systems than the Hamiltonian spectrum. For example, in \cite{Parker2018AHypothesis} a conjecture was put forward relating the Lanczos spectrum and dynamical aspects of operator growth. Concretely, linear growth of the low-order Lanczos coefficents -- the so-called edge of the spectrum -- was related to exponential growth of the ``Krylov complexity'' of the operator, to be defined below. A further important question is where the universality of chaotic behavior shows itself in the Lanczos spectrum. Here, based on results in \cite{Balasubramanian:2022dnj,SpreadC}, we propose that the Lanczos spectrum of a chaotic theory has a local mean and covariance that matches those of RMTs with the same density of states.
The main objective of the present article is to support and develop this conjecture  which complements the classic characterization of quantum chaos in terms of energy level statistics \cite{osti_4801180,PhysRevLett.52.1}.

Further interest in this tridiagonal approach to quantum chaos comes from problems in quantum gravity. One such problem is the black hole information paradox \cite{Hawking:1976ra}. The relation between this paradox and the physics of long times in chaotic systems was laid out in Ref. \cite{Maldacena:2001kr}: thermal relaxation in a unitary chaotic system with finite entropy should not go on forever; rather, at late times the magnitude of correlation functions should not decay to zero, and should instead plateau at a value exponentially small in the entropy of the system.
The Lanczos context sheds new light on this problem, since in this description of quantum dynamics, saturation of thermal relaxation is controlled by the tail of the Lanczos spectrum, i.e. the high-order Lanczos coefficients.
Relatedly, consider the ``bag-of-gold'' problem: the black hole interior, which grows indefinitely with time, can naively accommodate an infinite number of local degrees of freedom, in tension with the finite entropy associated to the horizon by Bekenstein and Hawking. A modern incarnation of this tension involves the proposal by \cite{Stanford:2014jda} that the black hole interior volume is related to the complexity of the underlying quantum state. In systems with finite entropy, the circuit complexity of states is upper bounded by the dimension of the Hilbert space. The  conjecture of \cite{Stanford:2014jda} then suggests that the interior volume of a black hole has an upper bound that is related to the black hole entropy. Recent work has proposed a resolution to both these puzzles \cite{Balasubramanian:2022gmo} based on non-perturbative effects in quantum gravity. The Lanczos approach also provides a useful perspective. First, it provides a sensible and computable notion of complexity of quantum operator/state evolution \cite{Parker2018AHypothesis,SpreadC}, which seems to be related to spacetime volume, at least in some simple situations \cite{Lin:2022rbf,Rabinovici:2023yex,Balasubramanian:2022gmo}. Second, in this approach both problems, late time saturation of thermal relaxation and of complexity, boil down to the decay of the tail of the Lanczos spectrum, as we will discuss below. The study of long times for this notion of complexity has appeared in \cite{Rabinovici:2020ryf,SpreadC,Balasubramanian:2022dnj,Erdmenger:2023shk,Bhattacharyya:2020qtd,Hashimoto:2023swv,Camargo:2023eev}. 

We start in Sec.~\ref{RMT_trid} by reviewing the results from \cite{Balasubramanian:2022dnj}, concerning the statistics of the Lanczos spectrum in RMT. Then, in Sec.~\ref{SecIII} we conjecture that quantum chaotic systems can be characterized by the universal fine-grained statistics of the Lanczos spectrum, that these statistics are well described by RMT, and that these statistics summarize the properties relevant to capture the late-time dynamics of chaotic systems. We support this conjecture in two ways. First, we study the Lanczos spectrum in examples of chaotic and integrable systems.  Second, we will show that the local mean and covariance of the Lanczos spectrum  control the dynamics of thermal relaxation, the survival amplitude, and spread complexity. 

Next, in  Sec.~\ref{SecIV}, for states with continuous support in the energy basis, we derive a formula for the long-time averaged probabilities of the Krylov basis states.  These quantities are generalized survival amplitudes, and our formula shows that they are completely determined by the density of states.  As a result, these late-time quantities, for initial states with continuous energy support, cannot discriminate between chaotic and integrable theories with the same density of states, even if the spectral correlations differ.  Our analysis then suggests a notion of energy eigenstate complexity, the average position of an eigenstate in the Krylov basis, relative to both Haar random initial states and initial states with continuous support in the energy basis.  We compute this notion for RMTs, and for several examples of chaotic and integrable Hamiltonian systems.  The fluctuations of this quantity across eigenstates with similar energies discriminates between integrable and chaotic systems, for initial states with continuous support in the energy basis. Anderson localization in the Lanczos chain leads to high variance in the eigenstate complexity because in order for a collection of more-localized eigenstates to cover the same probability distribution over the Lanczos chain, their individual average positions need to be spread out further apart. Localization also correlates the overlap between the initial state and an eigenstate with the average position of the same eigenstate along the chain.  This reduces the plateau values for random initial states compared to the plateaus for initial states with continuous support in the energy basis.

Lastly, in Sec.~\ref{SecV} we expand on the relation 
 between the long time behavior of spread complexity and  universality classes of quantum chaotic systems developed in
\cite{SpreadC,Erdmenger:2023shk}. We accomplish this by studying generalized Gaussian beta ensembles \cite{Dumitriu_2002}, labeled by a continuous Dyson index. We  end with a discussion in Sec.~\ref{SecC}.

\section{Random Matrix Theory (RMT) and the Lanczos approach}
\label{RMT_trid}

Here we describe the  framework of the Lanczos approach to Random Matrix Theory developed in
\cite{SpreadC,Balasubramanian:2022dnj}. 
Readers familiar with the material can skip this section.

\subsection{The Lanczos approach to quantum dynamics}\label{LQM}
In quantum mechanics the time evolution of a  state $\vert \psi (t)\rangle$ satisfies the Schr\"{o}dinger equation
\be
i\,\partial_t\,\vert \psi (t)\rangle=\,H\,\vert \psi (t)\rangle \; ,
\label{eq:se}
\ee
where $H$ is the Hamiltonian. Taylor expanding the solution one obtains
\be
\vert \psi (t)\rangle= e^{-iHt}\,\vert \psi (0)\rangle=\sum^\infty_{n=0}\frac{(-it)^n}{n!}\,\vert \psi_n\rangle\,,
\ee
where 
\be 
\vert \psi_n\rangle\equiv H^n\,\vert \psi(0) \rangle \;.
\ee 
The ordered list of vectors $ \vert \psi_n\rangle $ provides an optimized set in which to expand the evolving state \cite{SpreadC}. By applying the Gram–Schmidt procedure to the $\vert\psi_n\rangle $ we generate an ordered, orthonormal basis
$\mathcal{K}=\set{\ket{K_n}: n=0,1,2,\cdots}$. This is known as the Krylov basis. In this basis the Hamiltonian is tridiagonal
\be\label{triH}
H= \begin{pmatrix}
a_0 & b_1 &  & & & \\
b_1 & a_1 & b_2 & & &\\
& b_2 & a_2 & b_3 & & \\
& & \ddots & \ddots & \ddots & \\
& & & b_{N-2} & a_{N-2} & b_{N-1}\\
&  & &  &b_{N-1} & a_{N-1}
\end{pmatrix}\;,
\ee
The sequences $a_n,b_n$ are known as the Lanczos coefficients or the Lanczos spectrum. This tridiagonal form is also known as a ``Hessenberg form'' of a hermitian matrix.\footnote{The Hessenberg form is not unique, and is determined by the choice of the first vector in the new basis.  See the detailed discussion in \cite{SpreadC,Balasubramanian:2022dnj}.}  Conveniently, there are numerically stable algorithms that use Householder reflections to compute the Hessenberg form of a finite dimensional matrix ; see for example libraries like \cite{2020SciPy-NMeth,LAPACK-Hessenberg} or Mathematica.

The Lanczos spectrum can also be found from the moments of the Hamiltonian in the initial state
\be
\mu_n =
\langle K_0 |\, (iH)^n \,| K_0 \rangle \,,
\label{eq:momgenfn}
\ee
which in turn can be obtained from the survival amplitude, namely the amplitude for the state to remain unchanged \cite{SpreadC}. This is useful because the moment method remains valid for infinite-dimensional systems. To obtain the Lanczos coefficients from the moments, we use the fact that $\mu_n$ turns out to be a polynomial of the first $n$ Lanczos coefficients. This gives a recursion relation that can be solved efficiently through dynamic programming as outlined in \cite{SpreadC}, see also \cite{viswanath2008recursion}.

In the Krylov basis, the Hamiltonian can be pictured as a one-dimensional chain with nearest-neighbor hopping parameters given by  $b_n$, and with the initial state being the first site on the chain. It is then natural to analyze the average position along the chain as time evolves,
\be 
C(t) = C_\mathcal{K}(t) = \sum_{n} n \,\vert \braket{K_n}{\psi(t)}\vert^2 =
\sum_{n} n \,p_n(t)\;.\label{eq:spread_comp_def}
\ee
A related notion is the effective dimension of the Hilbert space explored by the time evolution, namely
\be 
C_{{\rm dim}}
= e^{H_\textrm{Shannon}}=e^{-\sum\limits_n p_n\log p_n}\;.
\label{ecom2}
\ee
Ref.~\cite{SpreadC} proves that these notions of spread of the wavefunction in the Krylov basis actually minimize the spread of the wavefunction in any possible basis at short times, making them natural quantifications of quantum complexity dubbed ``spread complexity''. 
As described above the wavefunction in the Krylov basis is a functional of the Lanczos coefficients  which are in turn determined  by the survival amplitude.  Thus, saturation of the survival amplitude will also imply saturation of the spread complexity.

Examples in which the computation of the Lanczos spectrum (i.e. the Hamiltonian in  tridiagonal form \eqref{triH} can be carried out analytically are sparse in the literature; see \cite{viswanath2008recursion,doi:10.1063/1.533010,PhysRevD.47.1640} for older examples and \cite{Parker2018AHypothesis,Caputa:2021sib,SpreadC,Muck:2022xfc,Balasubramanian:2022dnj,Caputa:2023vyr} for more recent ones.

\subsection{Chaos and Random Matrix Theory}

We are interested in applying the Lanczos approach described above to chaotic systems.
The Bohigas-Giannoni-Schmit conjecture \cite{PhysRevLett.52.1} states that the fine-grained features of the spectrum of a quantum chaotic Hamiltonian  are well approximated by the statistics of random matrices \cite{osti_4801180,PhysRevLett.52.1}.  In this section we will review the relevant features of Random Matrix Theory (RMT); see \cite{Meh2004,Guhr:1997ve,bookhaake,akemann2011oxford} for extensive accounts. An RMT can be defined as an ensemble of $N\times N$ matrices $H_{ij}$ with a distribution controlled by a potential $V(H)$
\be 
p(H)=\frac{1}{Z_{\beta_D}}e^{-\frac{\beta_D N}{4} \,\textrm{Tr}(V(H))}\;.
\ee
These matrices are self-adjoint and their entries can be either real, complex, or quaternions, corresponding to Dyson indices $\beta_D=1$, $2$, or $4$ respectively. Their eigenvalues are distributed according to the joint probability distribution
\be \label{jointb}
p(\lambda_1,\cdots ,\lambda_n)=Z_{\beta_D,N}\, e^{-\frac{\beta_D\,N}{4}\sum_{k}V(\lambda_k)}\,\prod_{i<j}\vert \lambda_i-\lambda_j\vert^{\beta_D}\;,
\ee
which follows from a change of variables to the diagonal form. This form also allows us to generalize $\beta_D$ to any positive value.

For Gaussian potentials Dumitriu and Edelman showed in a seminal work \cite{Dumitriu_2002} that the same joint probability distribution \eqref{jointb} arises from certain tridiagonal matrix ensembles. From a mathematical standpoint, a natural question that arises from that work is whether tridiagonal matrix ensembles can be found for random matrices distributed according to arbitrary potentials.  Such a tri-diagonal ensemble 
could be regarded as an ensemble of Hamiltonians in their Lanczos representation.  Below, we describe progress towards this goal that was recently achieved in \cite{Balasubramanian:2022dnj}.

\subsection{Tridiagonalizing Random Matrix Theory}
We now use the Lanczos approach to study RMT. More precisely, given a Hamiltonian $H$ and an initial quantum state $\vert\psi\rangle$, we can study aspects of quantum chaos and long times in the Krylov basis by applying the Lanczos algorithm and obtaining a tridiagonal Hamiltonian as in \eqref{triH}. In RMT we start with an ensemble of Hamiltonians, and consequently the Lanczos algorithm will produce an ensemble of tridiagonal random matrices. Given a particular RMT, defined by a potential $V(H)$ and/or its average density of states $\rho(E)$, the first goal is to find the statistics of these tridiagonal random matrices, namely the statistics of the Lanczos spectrum.

This goal was partly accomplished in Ref.~\cite{Balasubramanian:2022dnj}, where the average and covariance of the Lanczos spectrum were found. We now briefly review those results. Consider Hamiltonians whose densities of states are unimodal.  For such theories, numerical studies show that the Lanczos coefficients have a continuous large $N$ limit for initial states with sufficiently broad support on the energy eigenstates \cite{SpreadC,Balasubramanian:2022dnj}. By this we mean that, defining  $x\equiv n/N$,  $a(x) \equiv a_{xN}$ and $b(x) \equiv b_{xN}$ are continuous functions of $x$ as $N\to\infty$.
Assuming this continuity, \cite{Balasubramanian:2022dnj} derived a relation between the density of states and the Lanczos coefficients, \be \label{intdl}
\rho(E) =  \int_0^1 dx\, \frac{H(4\,b(x)^2-(E-a(x))^2)}{\pi\, \sqrt{4\,b(x)^2-(E-a(x))^2}}\;.
\ee
This formula is not limited to  random matrix theories; this equation comes from an analysis of the density of states of a tridiagonal matrix, and will be valid for the Lanczos coefficients of any system in the thermodynamic limit.
In this formula, the edge, i.e., the Lanczos coefficients for $n\sim\mathcal{O}(1)$, makes a vanishingly small contribution to the integral. So if we obtain $a(x)$ and $b(x)$ via this equation, we will lose information about the edge. This has consequences for some of our results in  Sec.~\ref{SecIII}, where we will need to re-introduce information about the edge to accurately describe some properties of the system.

While \eqref{intdl} works for any Hamiltonian, we can get more information assuming we have an RMT with potential $V(H)$. This was worked out in \cite{Balasubramanian:2022dnj}, making use of the fact that the Jacobian of the coordinate transformation from matrices to their tridiagonal form reads
\be 
J\propto  \prod_{n=1}^{N-1} b_n^{(N-n)\beta_D-1}\;,\label{eq:Jacobian}
\ee
so that the joint probability distribution is given by
\be
    p(a_0,\cdots ,a_{N-1},b_1,\cdots ,b_{N-1}) \propto \gr{\prod_{n=1}^{N-1} b_n^{(N-n)\beta_D-1}}\,e^{-\frac{\beta_D N}{4}\textrm{Tr}(V(H))}\;.\label{eq:joint_prob_dist}
\ee
In this expression  the exponents scale with $N$. So at large $N$, we can apply a saddle-point approximation to find the mean and covariance of the Lanczos spectrum. The details involving  various combinatorial identities are worked out in \cite{Balasubramanian:2022dnj}. Solving for the extrema of the potential produces the  equations\footnote{These equations are consistent with those derived in Ref.~\cite{doi:10.1063/1.533010,PhysRevD.47.1640}. Here they were derived using conventional saddle point techniques in the context of RMT.}
\bea\label{lrmt1}
    4\,(1-x) &=& b(x)\, \frac{\dx}{\dx b(x)}\,\gr{\int dE \,\frac{V(E)}{\pi \sqrt{4b(x)^2 - (E-a(x))^2}}}\;,\nonumber\\
    0&=&\frac{\dx}{\dx a(x)}\,\gr{\int dE \,\frac{V(E)}{\pi \sqrt{4b(x)^2 - (E-a(x))^2}}}\;.\label{eq:one_point_final}
\eea
Since the method of obtaining the potential $V(H)$ from the density of states $\rho(E)$ and vice-versa is already known, \eqref{eq:one_point_final} can be seen as another method of solving the integral equation \eqref{intdl}. For polynomial potentials the saddle point equations can be simplified; see \cite{Balasubramanian:2022dnj} for the explicit expressions.

The covariance of the Lanczos coefficients can also be obtained from the second derivatives of the potential. This calculation was carried out in \cite{Balasubramanian:2022dnj} with the outcome that the covariance between two nearby Lanczos coefficients away from the edge at large $N$ can be analytically approximated by
\bea\label{cov}
\text{cov} (a_i,a_{i+\delta}) \approx 4\,\text{cov} (b_i,b_{i+\delta}) \approx \frac{1}{2\pi\beta_D N} \int_0^{2\pi}\frac{e^{2ik\delta}dk}{\lambda(k, a_i, b_i)}\;,\nonumber\\
2\,\text{cov} (a_i,b_{i+\delta}) \approx \frac{1}{2\pi\beta_D N} \int_0^{2\pi}\frac{e^{ik(2\delta-1)}dk}{\lambda(k, a_i, b_i)}\;,\eea
where
\bea\label{lrmt2}
    \lambda(k, a_i, b_i) =\int dE~\frac{V'(E)-V'(a_i)}{b_i(E-a_i)}\,\eta\gr{\frac{E-a_i}{b_i},e^{ik}}\;,\label{eq:eig_defs}
\eea
and 
\bea
    \eta(x,t)=\frac{x}{\pi\sqrt{4-x^2}}\frac{1}{t+\frac{1}{t}-x}\;.
\eea
Notice that \eqref{cov} implies that the covariances of the $a_n$'s are 4 times the covariances of the $b_n$'s. We will use this below.

These analytical predictions were successfully compared with numerical computations for the following  potentials
\be\label{pot}
V_g(E)\equiv E^2, \,\,\,\,\, 
V_s(E)\equiv 3E^2-E^4+\frac{2}{15}E^6, \,\,\,\,\, 
V_q(E)\equiv \frac{1}{6}E^4-\frac{4}{9}E^3+\frac{8}{3}E\;.
\ee
For conciseness, we will use these same examples below. Expressions for the relevant densities of states and the mean and covariance for $a(x)$ and $b(x)$ can be found in \cite{Balasubramanian:2022dnj}.

\section{Universal statistics of the Lanczos spectrum}\label{SecIII}

\begin{figure}
    \centering
    \includegraphics[height=6cm]{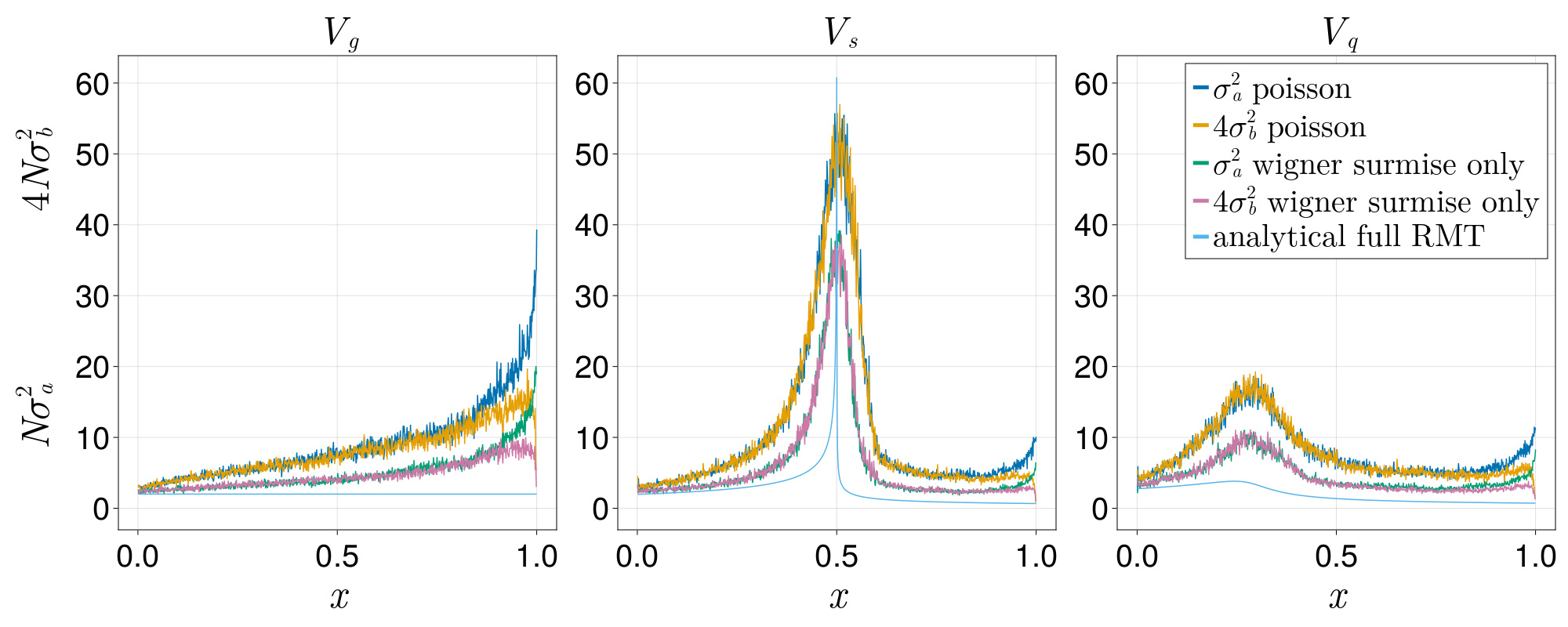}
    \caption{Variance of the Lanczos coefficients for random initial states for a Hamiltonian with Poisson distributed eigenvalues (solid blue for variance in $a_n$, orange for $4$ times variance in $b_n$), and for Hamiltonians chosen with Wigner-surmise-distributed \eqref{wign_surm} spacing (green for variance in $a_n$, purple for $4$ times variance in $b_n$), compared to analytical predictions of the corresponding values for random matrices with the same density of states (light blue), with Dyson index $\beta_D=1$. These plots show that the Wigner surmise alone is not sufficient to produce the Lanczos spectrum of a RMT.
    }\label{varlz_poisson}
\end{figure}

Quantum chaotic systems are famously conjectured to display a Hamiltonian spectrum whose fine-grained features are well described by RMT \cite{osti_4801180,PhysRevLett.52.1}.  By this we mean that in these systems the distribution of energy eigenvalues and their spacing is consistent with that of a random matrix theory, including the effect of eigenvalue repulsion seen in the latter. This conjecture has almost turned into a definition of quantum chaos \cite{akemann2011oxford,Guhr:1997ve}.  
Meanwhile, Lanczos and Krylov subspace methods have gained increasing attention as tools for analyzing the dynamics of time evolution in quantum mechanics, including chaotic systems.  
Here, we conjecture that 
\begin{center}
\emph{Quantum chaotic systems display a Lanczos spectrum well described by RMT.}
\end{center}
Note that a distribution of energy eigenvalues that exhibits gaps distributed according to the Wigner surmise (like in RMT) is not sufficient to produce the Lanczos spectrum of a RMT. For example, we selected eigenvalues by successively picking numbers following an unscaled Wigner surmise
\be
p(x_{i+1}-x_{i})\propto (x_{i+1}-x_{i})e^{-(x_{i+1}-x_{i})},\label{wign_surm}
\ee
scaled them down to lie between $0$ and $1$, and finally applied an inverse cumulative distribution function of the density of states to obtain eigenvalues whose nearest neighbor distributions match the proper Wigner surmise \cite{Meh2004, Guhr:1997ve}. In Fig.~\ref{varlz_poisson}, we plot the variance of the resulting Lanczos spectrum as well as those for Poisson distributed eigenvalues, and the analytical prediction for random matrices. The numerics demonstrate that the variance of the Lanczos spectrum in random matrices contains more information than the Wigner surmise.

In this section we support our conjecture and demonstrate its power. 
We first develop a method of estimating the density of states, and the average and covariances of the Lanczos spectrum of fixed integrable and chaotic Hamiltonians. We then compare the estimated variances with analytical predictions from RMT, and verify that they match for chaotic systems but differ wildly for integrable ones. Next, we show that the average and covariance of the Lanczos spectrum already contains the necessary information to accurately reproduce survival amplitudes and probabilities for random initial states. Combining this with the fact that the average and covariance of Lanczos spectrum can also reproduce the spectral form factor \cite{Balasubramanian:2022dnj}, our results demonstrate the power of our conjecture.

\subsection{Integrability and Chaos in the Lanczos Spectrum}\label{trid_predicts_chaos}

We will consider the dynamics of quantum systems that are time-reversal symmetric. This implies their state vectors are real. We want to start from a random initial state, by which we mean a state drawn from a distribution that is fixed under actions of orthogonal matrices.\footnote{If the systems were not time-reversal symmetric, then we would need to draw states from a distribution that is fixed under actions of unitary matrices instead.}  We will compute the Lanczos spectrum of several integrable and chaotic spin chains starting from such random initial states, and show that 
both chaotic and integrable systems match the analytical prediction \eqref{eq:one_point_final} of the average Lanczos spectrum of random matrix ensembles with a matching density of states. On the other hand, only chaotic systems without conserved quantities match the predicted variance \eqref{cov} of the RMT Lanczos spectrum. 

In what follows, we will compute the Lanczos spectrum of each theory for a single random initial state and compare it with RMT predictions.  The Hilbert spaces we work with are sufficiently high dimensional that it is exceedingly unlikely that the random state we pick happens to be in a ``special'' subspace in some way, e.g., a low-dimensional linear combination of energy eigenstates.  Even if the support of some energy eigenstates is exponentially small, the process of obtaining the Lanczos coefficients will amplify the effect of that eigenstate so that the dynamics will explore the full Hilbert space. 

We will examine the following spin chains
\bea
    H_\text{int1}&=&A_\text{int1}\left(\sum_{i=1}^{K-1} X_iX_{i+1} + \sum_{i=1}^K C_1 Z_i\right)\;,\nonumber\\
    H_\text{cha1}&=&A_\text{cha1}\left(\sum_{i=1}^{K-1} X_iX_{i+1} + \sum_{i=1}^K C_2 Z_i +C_3 X_i + H_\text{site}\right)+B_\text{cha1}\;,\nonumber\\
    H_\text{int2}&=&A_\text{int2}\left(\sum_{i=1}^{K-1} X_iX_{i+1}+Y_iY_{i+1}+C_4 Z_iZ_{i+1}\right)+B_\text{int2}\;,\nonumber\\
    H_\text{cha2}&=&A_\text{cha2}\left(\sum_{i=1}^{K-1} X_iX_{i+1}+Y_iY_{i+1}+C_5 Z_iZ_{i+1} + \sum_{i=1}^{K-2} C_6 Z_iZ_{i+2}  +  H_\text{site}\right)+B_\text{cha2}\;,\,\,\,\,\label{spin_chain_ham}
\eea
where $X_i$, $Y_i$, $Z_i$ are Pauli spin operators on site $i$; $A_k$, $B_k$ and $C_k$ are numerical parameters, and  $H_\text{site}=0.5Z_3+0.3X_3$ is a one-site disorder operator that helps breaks parity and symmetry under the action of $\sum_i Z_i$. The models defined by $H_\text{int1}$ and $H_\text{int2}$ are integrable, while $H_\text{cha1}$ and $H_\text{cha2}$ are chaotic as suggested by their level statistics and have been explored in the literature \cite{Ba_uls_2011, Rabinovici:2022beu}. For numerical computations, we set $C_1=C_2=C_4=1.1$, $C_3=-0.3$, $C_5=C_6=0.8$, and  simulated $K=12$ spins which have a $2^{12} = 4096$ dimensional Hilbert space. For simplicity, we have linearly scaled the Hamiltonians so that their eigenvalues lie between $-2$ and $2$. This eases the comparison with RMT. This rescaling is accomplished by setting $A_\text{int1}=0.1263$, $A_\text{cha1}=0.1152$, $A_\text{int2}=0.1198$, $A_\text{cha2}=0.1102$, and  $B_\text{cha1}=-0.1104$, $B_\text{int2}=0.5490$, $B_\text{cha2}=0.1092$.

If a Hamiltonian has degenerate eigenvalues with eigenvectors $\ket{E_i}$, dynamics starting from an initial vector will only explore a one-dimensional section of the space spanned by $\ket{E_i}$. Hence we can merge the degenerate eigenvectors into one basis element given by the projection of the initial vector onto the subspace. The initial vector in the new merged basis element will have an amplitude of $\sqrt{\sum_i \abs{\braket{\psi}{E_i}}^2}$. Doing this is necessary for the numerical calculations because otherwise floating point errors will break apart the degeneracy and the narrow gaps between the resulting eigenvalues will cause numerical instability. In the above list of Hamiltonians, $H_\text{int2}$ has some degenerate eigenvalues; we merge those as well as the corresponding directions in the initial state to obtain the Lanczos spectrum. This reduces the dimension of the Hilbert space in this case to  $N=2510$.

The Hamiltonians we consider will have Lanczos spectra which fluctuate around some coarse-grained local mean.  We can find this mean by locally smoothing the spectrum. We use SciPy \cite{2020SciPy-NMeth} to do this with a Savitzky-Golay filter \cite{savitzky1964smoothing} of order $1$ and window size $101$.  In order to estimate the variance around the local mean, we first subtract the raw Lanczos spectrum from its smoothed local mean, square the differences, and then apply another Savitsky-Golay filter with the same order and window size to obtain a smoothed estimate of the variance.   Note that explicitly averaging our results over randomly chosen initial states is not expected to match the analytical covariance in \eqref{cov}, as the covariance comes from randomness in both the eigenvalues of the Hamiltonian and the initial state, and is not conditioned upon a fixed Hamiltonian.

\begin{figure}
    \centering
    \includegraphics[height=6cm]{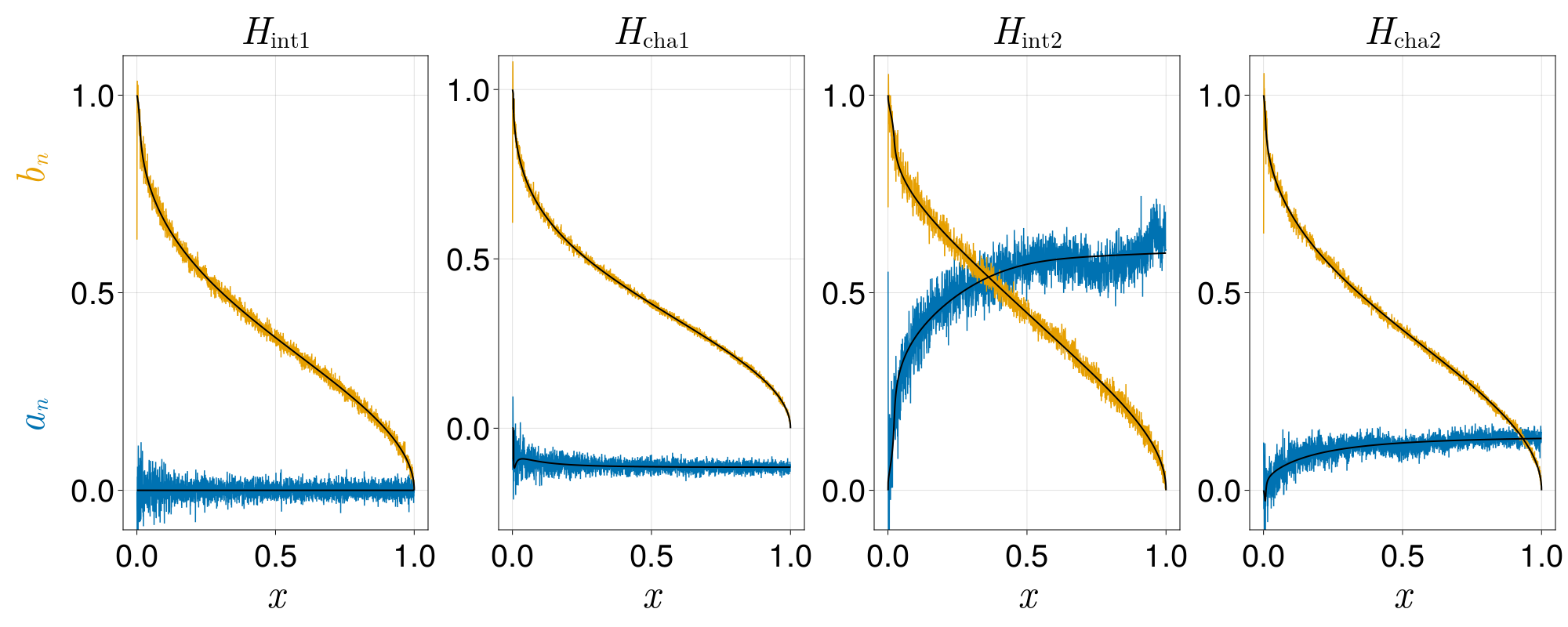}
    \caption{The analytical predictions (via 8th order Cheybyshev approximations of the density of states) of the mean of the Lanczos spectrum (blue for $a_n$, orange for $b_n$) from the density of states, compared to the actual values of the Lanczos spectrum (black), for the four Hamiltonians in \eqref{spin_chain_ham}. The matching is perfect for all models (integrable and chaotic).}\label{intcha_mean_val}
\end{figure}

To generate the analytical predictions of RMT for the mean and variance, we need to choose a potential $V(E)$ that gives an ensemble with the same density of states as the original theory. To do this, we first approximate the smooth density of states of the original theory as follows. Let $C_n(x)=U_n(x/2)$ be rescaled Cheybyshev $U$ polynomials.\footnote{The original Cheybyshev $U$ polynomials are orthogonal under the inner product $\langle f,g\rangle=\frac{2}{\pi}\int f(x)g(x)\sqrt{1-x^2}dx$ and the rescaling simply changes them to be orthogonal under $\langle f,g\rangle=\frac{1}{2\pi}\int f(x)g(x)\sqrt{4-x^2}dx$.} These polynomials can also be defined recursively via $C_0(x)=1$, $C_1(x)=x$, and $C_n(x)=xC_{n-1}(x)-C_{n-2}(x)$. Then, using the orthogonality and completeness of the Cheybyshev polynomials $\int C_n(x)C_m(x)\sqrt{4-x^2}dx=2\pi \delta_{mn}$, and writing $\rho(E)=\sqrt{4-E^2}\sum_i \alpha_iC_i(E)$, we have
\be
    \rho(E)=\sqrt{4-E^2}\sum_i \frac{C_i(E)}{2\pi}\int C_i(x)\rho(x)dx\approx \frac{\sqrt{4-E^2}}{2\pi N}\sum_i C_i(E)\sum_{\lambda} C_i(\lambda)\;.
\ee
Truncating the sum over $i$ gives a smooth controllable approximation to the density of states. We truncated our sum at $i=8$, at which point increasing the order of the Cheybyshev polynomials did not appreciably change our results.

\begin{figure}
    \centering
    \includegraphics[height=6cm]{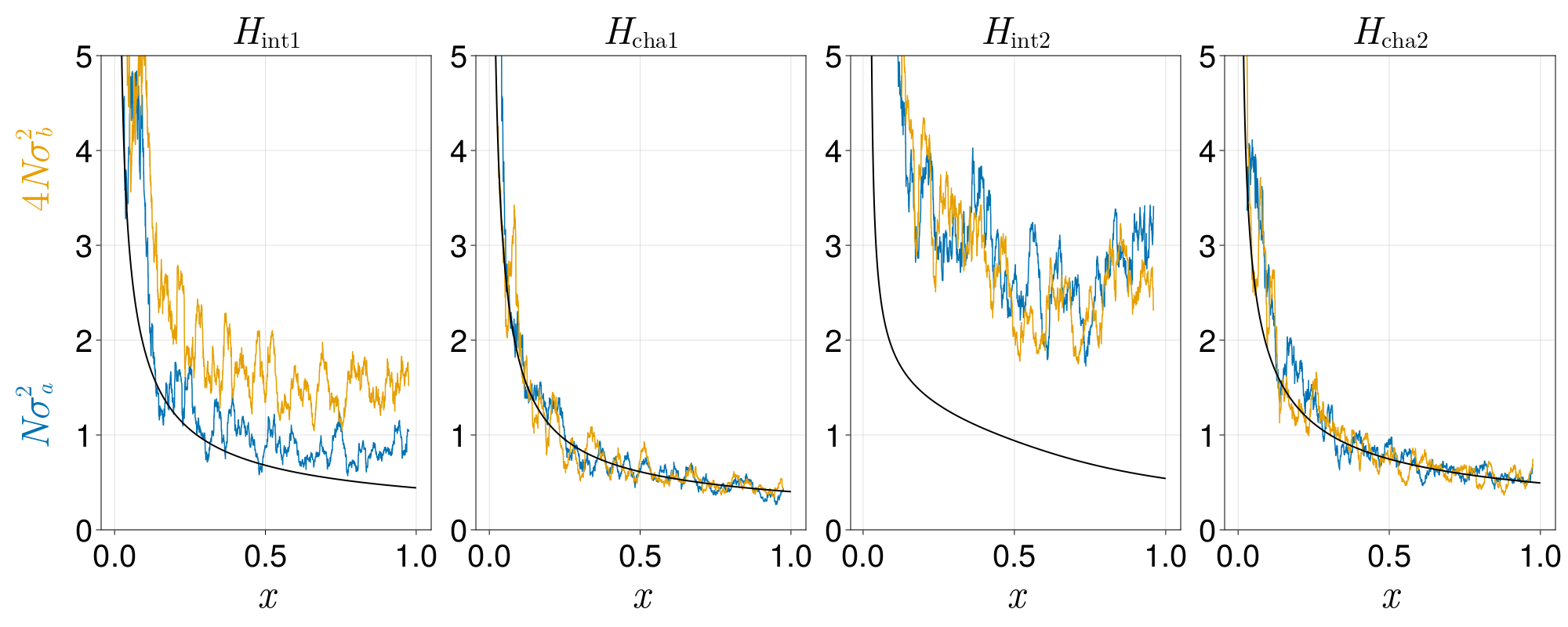}
    \caption{The analytical predictions (via 8th order Cheybyshev approximations of the density of states) of the variance of the Lanczos coefficients (black) compared to the numerically estimated variance of the Lanczos coefficients (blue for $a_n$, orange for $4$ times the variance of $b_n$), for the four Hamiltonians in \eqref{spin_chain_ham}. We see that realistic chaotic systems conform with random matrix theory predictions of the variance of the Lanczos spectrum while integrable systems do not.
    }\label{intcha_mean_variance}
\end{figure}

Next, following \cite{Balasubramanian:2022dnj}, we found the potential $V(E)$ using the saddle point equation of RMT, namely
\be 
\frac{1}{4}V'(\omega)= -\text{p.v.}\int dE\frac{\rho(E)}{\omega-E}\;,
\ee
where $\omega$ lies between $-2$ and $2$, and where p.v. means we want the principal value of the integral \cite{https://doi.org/10.48550/arxiv.1510.04430}. In particular, since $\text{p.v.}\int dE\frac{E^n\sqrt{4-E^2}}{\omega-E}=\pi\omega^{n+1}$ and our estimate for $\rho(E)$ is of the form $\text{poly}(E)\sqrt{4-E^2}$, this gives us a polynomial approximation for $V(E)$. We then inserted this $V(E)$ into (\ref{lrmt1}, \ref{lrmt2}) above to find the RMT predicted mean and variance of the Lanczos spectrum. All of the above Hamiltonians are real and belong in the $GOE$ class, so we take $\beta_D=1$.

Figs.~\ref{intcha_mean_val} and \ref{intcha_mean_variance} compare the analytical mean and variance predicted by RMT with the numerical estimate of the mean and variance obtained using the above smoothing methods. We find that our integrable and chaotic spin chains all match the predicted RMT mean Lanczos spectrum.  This matching was expected because the mean Lanczos spectrum can be derived from only the density of states as in \eqref{eq:one_point_final}.  In contrast, only the chaotic spin chains have variances that match the RMT predictions. Notice that the matching does not apply to the edge and tail values of the Lanczos coefficients, namely $a_n$ and $b_n$ for $n\sim O(1)$ and $N-n\sim O(1)$.  This breakdown occurs because the Savitzky-Golay smoothing cuts off, and the Cheybyshev approximation becomes inaccurate. In fact, the analytical predictions from RMT in both \eqref{eq:one_point_final} and \eqref{cov} are also inaccurate at the edge, due to the breakdown of the continuity assumption: at the edge the value of $b(x)$ in effect suddenly drops down from $b_1$ to zero \cite{Balasubramanian:2022dnj}. Thus, if the edge or tail values are important for some specific question, as they will be below, we will need to compute them separately.

\subsection{Universal statistics are enough to predict the long time dynamics}\label{trid_predicts_SFF}

We would like to determine what aspects of RMT spectral statistics and dynamics are described by 
the mean and covariance of the Lanczos spectrum. To explore this, we can ask whether other known quantities and averages of RMT can be reproduced just from the mean and covariance of the Lanczos spectrum. An example of this is given in \cite{Balasubramanian:2022dnj}, where it was shown that in RMT the analytical average and covariance of the Lanczos coefficients (\eqref{eq:one_point_final} and \eqref{cov}) relative to a random initial state are enough to reproduce the spectral form factor (SFF) across all time scales. The spectral form factor is given by
\be 
\textrm{SFF}=\frac{Z_{\beta-i\,t}\,Z^*_{\beta+it}}{Z_{\beta}^2}\;,
\ee
with $Z_\beta=\sum_i\,e^{-\beta\,E_i}$ being the partition function of the Hamiltonian, and $E_i$ the eigenvalues. This implies that the statistics of the Lanczos spectrum contain key aspects of the spectrum of the Hamiltonian in chaotic theories, such as the spectral rigidity controlling the ramp of the spectral form factor \cite{Guhr:1997ve,akemann2011oxford}. In addition, \cite{Balasubramanian:2022dnj} showed that the spread complexity of the Thermofield Double (TFD) state also matched for random matrix Hamiltonians and tridiagonal Hamiltonians produced  with the same mean and covariance as the Lanczos spectrum. This matching occurred even though the analytic formulae for the Lanczos coefficients \eqref{eq:one_point_final} and \eqref{cov} are inaccurate at the edge of the spectrum, as we explained earlier.  This is because computing the SFF and the TFD state involves a re-diagonalization of the tridiagonal Hamiltonian, a process which dilutes the error in the edge via mixing with the much larger central part of the Lanczos spectrum.

\begin{figure}
    \centering
    \includegraphics[height=6cm]{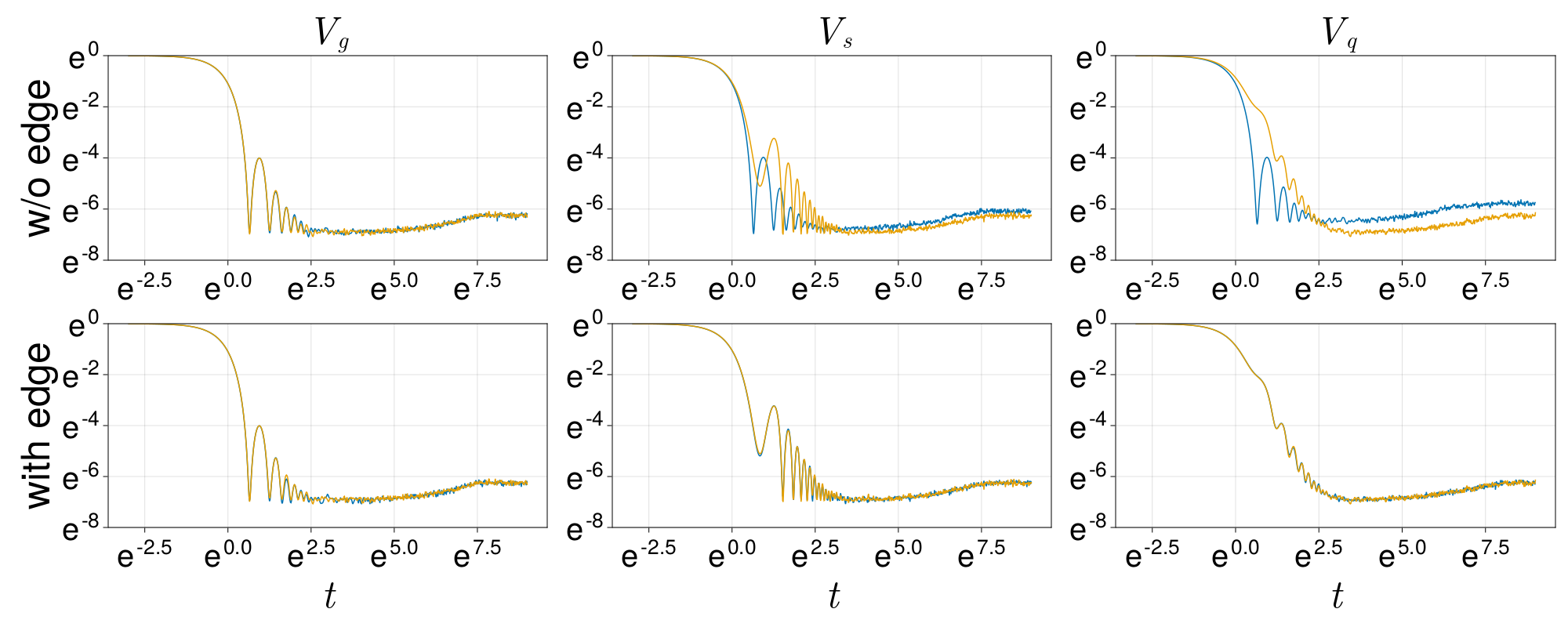}
    \caption{Survival amplitudes, averaged over $256$ instances of a random initial state evolved with an $N=1024$ random matrix drawn from a distribution with the labeled potentials with Dyson index $2$. The orange graphs are the values obtained from a stretched spectrum while the blue graphs are the values from tridiagonal matrices following the analytical mean and covariance. Top row is without the edge correction and bottom row is with the edge correction, changing the first $n=7$ Lanczos coefficients. This demonstrates the necessity and accuracy of the edge correction to our analytical formula.}
    \label{RISSA-bad}
\end{figure}

\begin{figure}
    \centering
    \includegraphics[height=6cm]{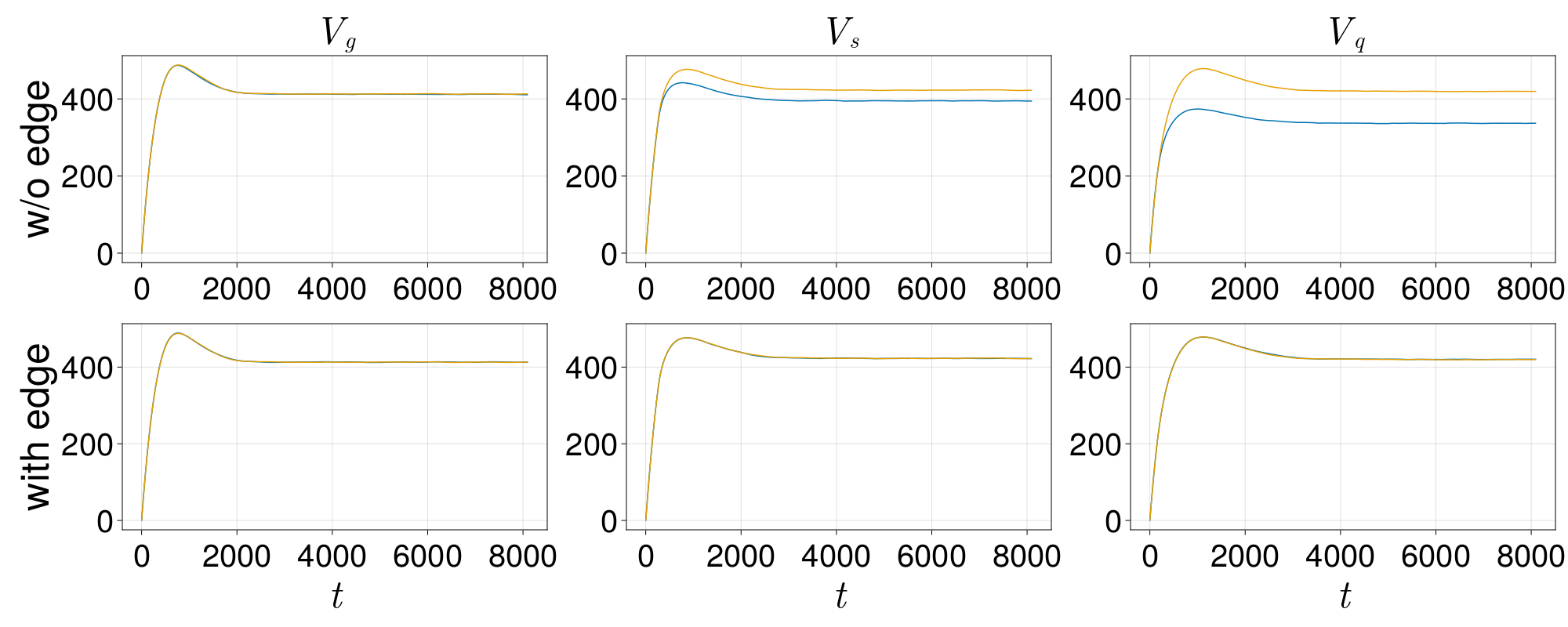}
    \caption{The spread complexity of a random initial state for an $N=1024$ random matrix with the potential  in \eqref{pot}, averaged over 256 samples. Results from a stretched spectrum (orange) and using the analytical mean and covariance of the Lanzcos coefficients (blue). Top is without the edge correction, bottom is with the edge correction up to $n=7$. This demonstrates the necessity and accuracy of the edge correction to our analytical formula.}
    \label{RISSC}
\end{figure}

If we try to compute the survival amplitudes of the canonical initial state which is localized on the first site of the Lanczos chain $\ket{K_0}$ of our approximate tridiagonal Hamiltonian, the survival probabilities $\abs{\braket{\psi(t)}{\psi(0)}}^2$ and spread complexities no longer match RMT because of the errors at the edge of the Lanczos spectrum (top rows of Figs.~\ref{RISSA-bad} and \ref{RISSC}). Here, we show that a small edge correction that adjusts the means of the first $O(1)$ Lanczos coefficients is sufficient to repair this inaccuracy.
First, to compute the exact survival amplitude and spread complexity of random initial states, we must generate RMT Hamiltonians. RMT Hamiltonians with general potentials can be generated accurately via Metropolis sampling \cite{hastings1970monte}, but this is very slow. A decent and much faster approximation takes the eigenvalues $E_i$ of a Gaussian random matrix and stretches the spectrum to match the density of states of the target non-Gaussian RMT. We can do this by applying on each energy eigenvalue the analytical cumulative distribution $f(E)=\frac{1}{2\pi}\int_{-2}^{E} dE'~\sqrt{4-E'^2}$ associated with the density of states of Gaussian RMT, followed by the inverse cumulative distribution $g^{-1}(E)=\int_{-\infty}^E dE'~\rho(E')$ of the density of states of the target non-Gaussian RMT. In this way we obtain an approximation $\tilde{E}_i=g(f(E_i))$ of the spectrum of the non-Gaussian Hamiltonian. In Fig.~(\ref{RISSA-bad}) this is done for the three different potentials in \eqref{pot}.

Next, we must sample tridiagonal matrices with the constraint that they match the required mean \eqref{eq:one_point_final} and covariance \eqref{cov}.
To sample a Gaussian distribution with a known covariance matrix $M^{-1}$, we first write the covariance matrix as a product $M^{-1}=LL^T$, a representation known as the Cholesky decomposition. We then sample i.i.d gaussian variables $\xi_i$ with covariance $\overline{\xi_i \xi_j}=\delta_{ij}$. Then, a linear transformation of these variables $x_i=\sum_j L_{ij}\xi_j$ has covariance\footnote{Our approximate $M^{-1}$  calculated by our methods above is not exactly symmetric since translation invariance isn't exact for finite $N$. One can symmetrize it by taking the average of $M^{-1}$ with its transpose, without affecting the results at large $N$.}
\be
\overline{x_i\,x_j}=L_{ik}\,L_{jl}\,\overline{\xi_k \,\xi_l}=L_{ik}\,L_{jk}=M^{-1}_{ij}\;.\label{x_cov}
\ee
Adding the $x_i$'s to our mean values will produce an ensemble with the desired mean and variance. Using this procedure to determine the Lanczos coefficients and computing the survival probability and spread complexity, we arrive at the results in the top rows of Fig.~\ref{RISSA-bad} and \ref{RISSC}.

The results of this first approximation deviate from the exact result because, as we discussed, our analytical formulae for the statistics of the Lanczos spectrum are inaccurate at the edge of the spectrum. To correct the error, we can estimate the low-order Lanczos coefficients more accurately from the moments of the Hamiltonian using the method described in Sec.~\ref{LQM}, near \eqref{eq:momgenfn}, instead of using the approximation from Eq.~\eqref{eq:one_point_final}.

The moments of the Hamiltonian in the random initial state $\ket{\psi}$ are given by $\bra{\psi}H^n\ket{\psi}=\sum_i E_i^n \abs{\braket{\psi}{E_i}}^2$. For small $n\sim O(1)$, the fact that $E_i^n$ is smooth causes the noise in $\abs{\braket{\psi}{E_i}}^2$ to average out. This implies that the moments may be approximated by the integral
\be
\bra{\psi}H^n\ket{\psi}=\sum_i E_i^n \abs{\braket{\psi}{E_i}}^2 \approx  \frac{1}{N}\sum_i E_i^n=\int dE~E^n\rho(E)\;,
\ee
in the large-$N$ limit.\footnote{In the large $N$ limit, random normalized vectors $\psi$ are approximately gaussian in every degree of freedom, and so $\beta_DN\abs{\braket{\psi}{E_i}}^2$ are approximately independent and distributed according to a $\chi^2_{\beta_D}$ distribution, with an average of $\beta_D$ and a variance of $2\beta_D$. Hence, the mean can be estimated as described above, and the total variance is approximately $\sum_i E_i^{2n} \frac{2}{\beta_D N^2}\approx \frac{2}{\beta_DN}\int dE~E^{2n} \rho(E)$.} Then, to estimate the Lanczos coefficients at the edge, we insert these moments into the algorithm briefly explained in Sec.~\ref{LQM} (see details in \cite{SpreadC} and \cite{viswanath2008recursion}). To obtain the full Lanczos spectrum, we replace the first $n=7$ mean values of the Lanczos coefficients obtained via \eqref{eq:one_point_final} with the edge correction from this method. We then repeat the procedure described above for adding fluctuations around the mean consistent with RMT statistics.\footnote{Note that we are here applying an approximation to the noise in the bulk of the spectrum at the edge. In fact, this is somewhat incorrect because the noise in the edge is dominated by the randomness of $\abs{\braket{\psi}{E_i}}^2$. Since this error is both localized to a small $O(1)$ portion of the chain and has a magnitude of $O(\sqrt{N})$, this does not have a big effect.}

In the lower rows of Figs.~\ref{RISSA-bad} and \ref{RISSC} we show the results with the edge correction, i.e., the survival amplitude for the evolution of the random initial state from the stretched spectrum, as well as the same quantity for the canonical initial state $\ket{K_0}$ evolved by a tridiagonal Hamiltonian with Lanczos statistics determined by \eqref{eq:one_point_final} and \eqref{cov}, with the edge correction.
The RMT results now match those of the tridiagonal approximation across all time scales. Thus, approximating random Hamiltonians with a tridiagonal approximation with the right mean and covariance is sufficent to produce the long time dynamics of RMTs, and hence, combining  with results of Sec.~\ref{trid_predicts_chaos}, of chaotic theories.

\section{Finding the plateau values}\label{SecIV}
At very long times in chaotic systems, the wavefunction in the Krylov basis reaches a stationary regime in which the probabilities associated with each basis vector $\abs{\braket{\psi(t)}{K_n}}^2$, although not constant in time, fluctuate around their long time average 
\be
\frac{\omega(x)}{N}=\overline{\abs{\braket{\psi(t)}{K_{xN}}}^2} \, .
\ee
For example, the survival probability 
$\abs{\braket{\psi(t)}{\psi(0)}}^2=\abs{\braket{\psi(t)}{K_0}}^2$ is the probability of being in the first state of the Krylov chain, 
and saturates at long times. For an initial state with constant support over the energy basis, if we average the probability of a basis element $\ket{K_n}$, 
\be
    \abs{\braket{K_n}{\psi(t)}}^2 = \bra{K_n}\frac{1}{N}\sum_{E,E'} \ket{E}\bra{E'}e^{-i(E-E')t} \ket{K_n}\;,\label{avg_over_phase}
\ee 
over long times, the $e^{-i(E-E')t}$ averages out to $0$ unless $E=E'$, so we get $\abs{\braket{K_n}{\psi(t)}}^2 = \frac{1}{N}$, i.e., $\omega(x)=1$. But in general, using the same trick of averaging over phases as in \eqref{avg_over_phase}, the long-time average probability distribution in the Krylov basis has an extra factor
\be
    \overline{\abs{\braket{\psi(t)}{K_{xN}}}^2} = \sum_{E} \vert\langle\psi\vert E\rangle\vert^2  \braket{K_{xN}}{E}\braket{E}{K_{xN}}. \label{phase_avg}
    \ee
Below, we explain how we can work this into an analytical form in the large-$N$ limit for states with continuous support over the energy basis.

Suppose we consider families of states and Hamiltonians parametrized by $N$ such that in the large $N$ limit the Hamiltonian has a fixed density of states $\rho(E)$ and the state has support $P(E)$ in the energy basis.  In this limit, for general initial states with continuous support over the energy basis, we find analytical expressions for the plateau values in terms of the local mean of the Lanczos spectrum.  By continuous support in the large $N$ limit we mean that there is a continuous function  $P(E)$ such that the distribution of the initial state over the energy basis satisfies
\be
N\vert\langle\psi\vert E\rangle\vert^2 \equiv P(E)\;.
\ee
This assumption does not hold for random initial states in the large $N$ limit, where our formula for the stationary probability will fail.
But it holds for example when considering the Thermo-Field Double (TFD) as the initial state. The TFD state is defined by
\be 
\vert\psi_{\beta}\rangle \equiv\frac{1}{\sqrt{Z_{\beta}}}\sum_n e^{-\frac{\beta E_n}{2}}\vert E_n\rangle \otimes\vert E_n\rangle\;,
\label{TFDdef}
\ee
in the tensor product of the original Hilbert space with itself,  where $\vert E_n \rangle$ are the eigenstates of the Hamiltonian with energies $E_n$. Evolving with, say, the right Hamiltonian one obtains $\vert\psi_{\beta} (t)\rangle =\vert\psi_{\beta+2it}\rangle$, and the survival amplitude is the analytically continued partition function. The function $P(E)=e^{-\beta E}/Z(\beta)$ is a smooth function of the energy.

With this assumption, we can estimate the plateau probability of each Krylov basis state \eqref{phase_avg} analytically by a method similar to the derivation of the integral equation relating the density of states to the average Lanczos spectrum \eqref{intdl} in \cite{Balasubramanian:2022dnj}.  Intuitively, one can argue that in the large $N$ limit the Lanczos spectrum becomes continuous, and we can break the 1-D chain into intervals whose sizes divided by $N$ become infinitesimal in the large $N$ limit, but remain much larger than $1$ site e.g., $dx=\frac{1}{\sqrt{N}}$. In each interval, the density of states is given by $\tilde{p}(x,E)=\frac{1}{\pi\sqrt{4b^2-(E-a)^2}}$, which we can interpret as a density of states over both $x$ and $E$. As we will derive below, this expression can essentially be integrated over $E$ to give the long time average of the probabilities over the Krylov chain.

\begin{figure}
    \centering
    \includegraphics[height=6cm]{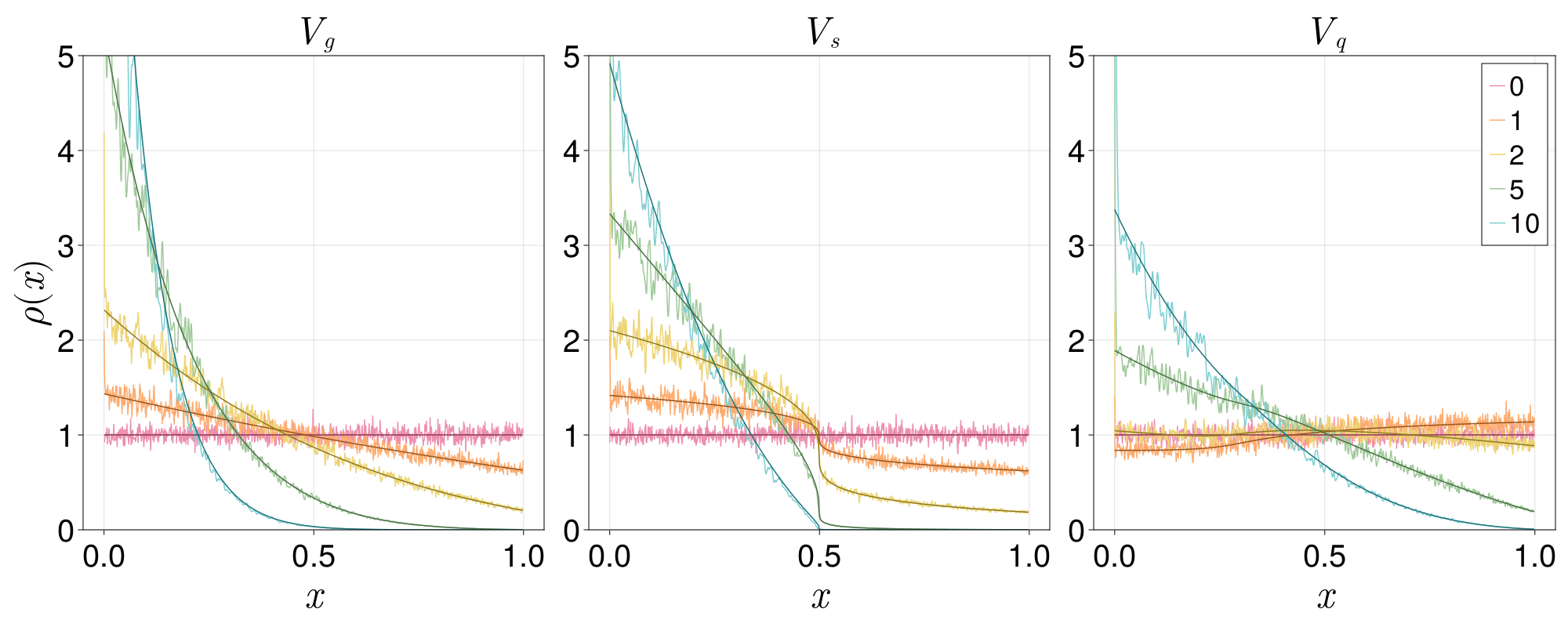}
    \caption{The long-time ($t=8N$) average probability distribution of a TFD state in its Krylov basis with $\beta=0,1,2,5,10$ for $N=1024$ random matrices averaged over $256$ draws of random matrices (noisy, lighter lines), compared to the analytical value obtained from \eqref{pstfd2} (smooth, darker lines). We see that that our analytical result for the late time distribution over the Krylov basis $\omega(x)$ is accurate.
    }\label{pstatio}
\end{figure}

In more detail, recall that in the Krylov basis, the dynamics is that of a single particle in a 1-D chain with hopping parameters given by the Lanczos coefficients $a_n$ and $b_n$. Consider then the problem of estimating the moments $\bra{K_i}H^k\ket{K_i}$. Since each application of the Hamiltonian can only shift our state by one site, the $H^k$ moment will only depend on the hopping amplitudes to and from sites within a distance $k$ of $i$. We gave evidence in \cite{Balasubramanian:2022dnj,SpreadC} that in the large $N$ limit the Lanczos coefficients converge to continuous functions $a(x)$ and $b(x)$ of $x\equiv n/N$. So when $k\ll N$, since our result is only affected by hopping amplitudes within a distance $k$ of state $i$, we may approximate $\bra{K_i}H^k\ket{K_i}\approx\bra{K_i}T(a_i,b_i)^k\ket{K_i}$, where $T(a,b)$ is an infinite tridiagonal matrix with constant diagonal with entries $a$ and constant off-diagonals with entries $b$, i.e.
\be\label{Triav}
    T(a,b)=\begin{pmatrix}\ddots&\ddots&&&\\
    \ddots&a&b&&\\
    &b&a&b&\\
    &&b&a&\ddots\\
    &&&\ddots&\ddots
    \end{pmatrix}.
\ee
From section 4.2 in \cite{Balasubramanian:2022dnj}, an argument based on the combinatorics of Dyck paths gives us that
\be
    \bra{K_i}T(a_i,b_i)^k\ket{K_i}=\int_{a_i-2b_i}^{a_i+2b_i} dE\,\frac{E^k}{\pi \sqrt{4b_i^2-(E-a_i)^2}}\;.
\ee
Then, for fixed degree polynomials $f(E)$, in the large $N$ limit we have
\be
    \bra{K_i}f(H)\ket{K_i}\approx \bra{K_i}f(T(a_i,b_i))\ket{K_i}= \int_{a_i-2b_i}^{a_i+2b_i} dE\,\frac{f(E)}{\pi \sqrt{4b_i^2-(E-a_i)^2}}\;.\label{krylov_density_gen}
\ee
Choosing $f(E)=P(E)$ finally gives us the approximation
\bea
    \omega(x) &=& N \overline{\abs{\braket{\psi(t)}{K_{xN}}}^2} = \sum_{E} P(E) \braket{K_{xN}}{E}\braket{E}{K_{xN}}=\bra{K_{xN}}P(H)\ket{K_{xN}} \nonumber\\&\approx & \int_{a(x)-2b(x)}^{a(x)+2b(x)} dE\,\frac{P(E)}{\pi \sqrt{4b(x)^2-(E-a(x))^2}}\;.\label{krylov_density}
\eea
Only in this last step does the continuity of $P(E)$ factor into our argument. Choosing, for example, a random initial state would give a $P(E)$ that cannot be estimated by fixed order polynomials as $N$ increases, and can therefore not fulfill the condition on $f(E)$.
For the time evolution of the TFD state we explicitly have
\be\label{pstfd2} 
\omega(x)= \frac{1}{Z(\beta)}\,N \int dE \,\frac{e^{-\beta E}}{\pi\,\sqrt{4b(x)^2-(E-a(x))^2}}= \frac{1}{Z(\beta)}\,N \,I_0(2\beta b(x))e^{-\beta a(x)}\;,
\ee
where $I_0$ is the Bessel $I_0$ function. This distribution should be interpreted as an average distribution at very long times. Equivalently, we take the exact distribution as a function of time and then average it over time. In Fig.~\ref{pstatio} we verify this prediction numerically by plotting the analytical formula for the plateau distribution of the time evolved TFD state \eqref{pstfd2}, along with an ensemble average of the exact probability distribution computed numerically by the methods in \cite{SpreadC}, showing a perfect match with RMTs with the three different potentials. Note that the method we have developed for evaluating the Krylov plateau distribution $\omega(x)$ works for any theory, and not just RMTs, in the large system limit, provided the initial state has continuous support in the energy basis. As a result, in this scenario with an initial state with smooth support in the energy basis, the plateau cannot discern between integrable and chaotic systems with the same density of states.

Once we have the stationary distribution, we can compute the plateau of the spread complexity by simply integrating $N\int_0^1 x \, \omega(x)dx$. We can also compute the Shannon entropy of the average probability distribution $\omega(x)$. Alternatively, we can compute the time average of the entropy of the Krylov state probability distribution, $H_\text{Shannon}$ defined in \eqref{ecom2}.  These two entropic quantities are related as follows. Consider the state $\ket{\psi(t)}$ at long times $t$. While the magnitude $\vert\langle\psi\vert E\rangle\vert^2$ is a continuous function $P(E)$, the complex phases of $\langle\psi\vert E\rangle$ for different $E$ can be approximated as independent random variables with mean zero at late times $t$, because the phases are rotating separately through Hamiltonian time evolution.\footnote{Strictly speaking this argument only holds for Hamiltonians whose spectra consist of incommensurate energies or at least energies which have large LCMs.  In practice for interacting theories this will be true for the majority of energies in the spectra.} Then, since $\langle\psi(t)\vert K_n\rangle=\sum_E \langle\psi(t)\vert E\rangle\langle E\vert K_n\rangle$, $\langle\psi\vert K_n\rangle$ is heuristically a sum of many independent random variables and is therefore in the large $N$ limit Gaussian distributed with mean zero in both its real and imaginary parts. Also since the phases of  $\langle\psi(t)\vert E\rangle\langle E\vert K_n\rangle$ are uniformly random, the variances of the real and imaginary components are the same since we have rotational invariance in the complex plane.  
Therefore, each of the variances (in the real and imaginary components) is given by half the expectation of $\vert\langle\psi\vert K_n\rangle\vert^2$, which is something we have already computed, namely $\omega(x)/N$. The distribution of $p_n=\vert\langle\psi\vert K_n\rangle\vert^2$ is then the sum of the squares of two independent Gaussian-distributed variables, which is just an exponentially decaying distribution. The difference in contribution to entropy between exponentially distributed $p_n$ and a constant $\bar{p}_n$ reads
\be
   -\bar{p}_n\ln\bar{p}_n+\int_0^\infty \frac{e^{-p_n/\bar{p}_n}}{\bar{p}_n}dp_n~p_n\ln p_n \approx \bar{p}_n(1-\gamma_E)\;,
\ee
as $\bar{p}_n\to 0$. For large $N$, $\bar{p}_n$ is small, so we may use this limit. Summing the contribution of every $p_n$ gives us that
\be
    H_\text{Shannon}\approx \gamma_E-1-\sum_n \bar{p}_n\ln \bar{p}_n\approx \ln N+\gamma_E-1-\int dx\, \omega(x)\ln\omega(x)\;,
\ee
which can be verified numerically as well. This Shannon entropy plateau can be used to provide a spread complexity version of the Page curve, see \cite{Caputa:2023vyr}.

Summarizing, for initial states with continuous support in the energy basis, the plateau can be found analytically in terms of the mean Lanczos spectrum. It therefore does not codify the chaotic nature of the system. The reason is that the relation between the mean Lanczos spectrum and density of states is universal, and applies as well to integrable systems. As long as the integrable and chaotic systems being compared have the same density of states (or equivalently, the same mean Lanczos spectrum) the plateau of Krylov/spread complexity will be the same. On the other hand, as we have seen in the previous section, the covariance of the Lanczos spectrum does differ between chaotic and integrable theories and shows up through the approach of the spread complexity to the plateau.

\subsection{Eigenstate Complexity}\label{SecVI}
While we have shown that the plateau arising from initial states with continuous support in the energy basis  depends only on the density of states, the same is not true for random initial states (i.e., Haar random rotations of an initial state) and in general for states whose support in the energy basis $P(E)$ is not continuous. In particular the plateau is lowered by  ``roughness'' in $P(E)$. Some intuition for this fact can be gleaned from looking at the location of the eigenstates along the chain. Given an eigenstate $\vert E\rangle$ of energy $E$, we define its complexity relative to $\ket{\phi}$ 
by its mean position on the Krylov basis associated with the time evolution of $\ket{\phi}$,
\be
    C_E=\langle E\vert\,\hat{n}\, \vert E\rangle\;,
\ee
where $\hat{n}$ is the position operator in the Krylov basis, namely
\be
\hat{n}\equiv \sum_{n}\,n\,\vert K_n\rangle\langle K_n\vert\;. 
\ee
We will use a bar above such quantities to denote the local mean over an energy band. For example, 
we will consider the local variance of the eigenstate complexity $C_E$, by which we mean  $\sigma_E^2=\overline{C_E^2} -\overline{C_E}^2$.\footnote{Notice that this is not the same thing as the variance of the operator $\avg{\hat{n}^2}-\avg{\hat{n}}^2$.} 

The result of \eqref{krylov_density_gen}, namely the expectation values of polynomials of the Hamiltonian in a Krylov state, holds for any initial state so long as the Lanczos spectrum remains continuous in the large $N$ limit.\footnote{Note that while \eqref{krylov_density} requires $P(E)$ to be a continuous function, \eqref{krylov_density_gen} only requires $f(E)$ to be a polynomial and does not require the Krylov basis to be generated from an initial state with continuous support over the energy basis.} We can use this fact to estimate the local mean of the eigenstate complexity in the large $N$ limit. Choose $f(E)$ to be a polynomial approximation of an indicator function on an interval $[E,E+dE]$ (or any other local weight function) that remains fixed in size as $N$ increases. By applying \eqref{krylov_density_gen}, we have that 
\be
    \sum_{E'\in [E,E+dE]}\braket{K_n}{E'}\braket{E'}{K_n}=\bra{K_n}f(H)\ket{K_n}\approx \frac{H(4b_n^2-(E-a_n)^2) }{\pi \sqrt{4b_n^2-(E-a_n)^2}}dE\;,\label{energy_slice_density}
\ee
and so the local mean of the eigenstate complexity, averaged across the interval $[E,E+dE]$ is given by
\bea
    \overline{C_E}dE&=&dE\sum_n \frac{H(4b_n^2-(E-a_n)^2) }{\pi \sqrt{4b_n^2-(E-a_n)^2}}
    \approx N dE\int dx~\frac{H(4b(x)^2-(E-a(x))^2)}{\pi \sqrt{4b(x)^2-(E-a(x))^2}}\;,\label{analytical_eig_comp}
\eea
where we have replaced the boundaries of the integral in \eqref{krylov_density_gen} with a Heaviside step function $H(x)$.

\begin{figure}
    \centering
    \includegraphics[height=6cm]{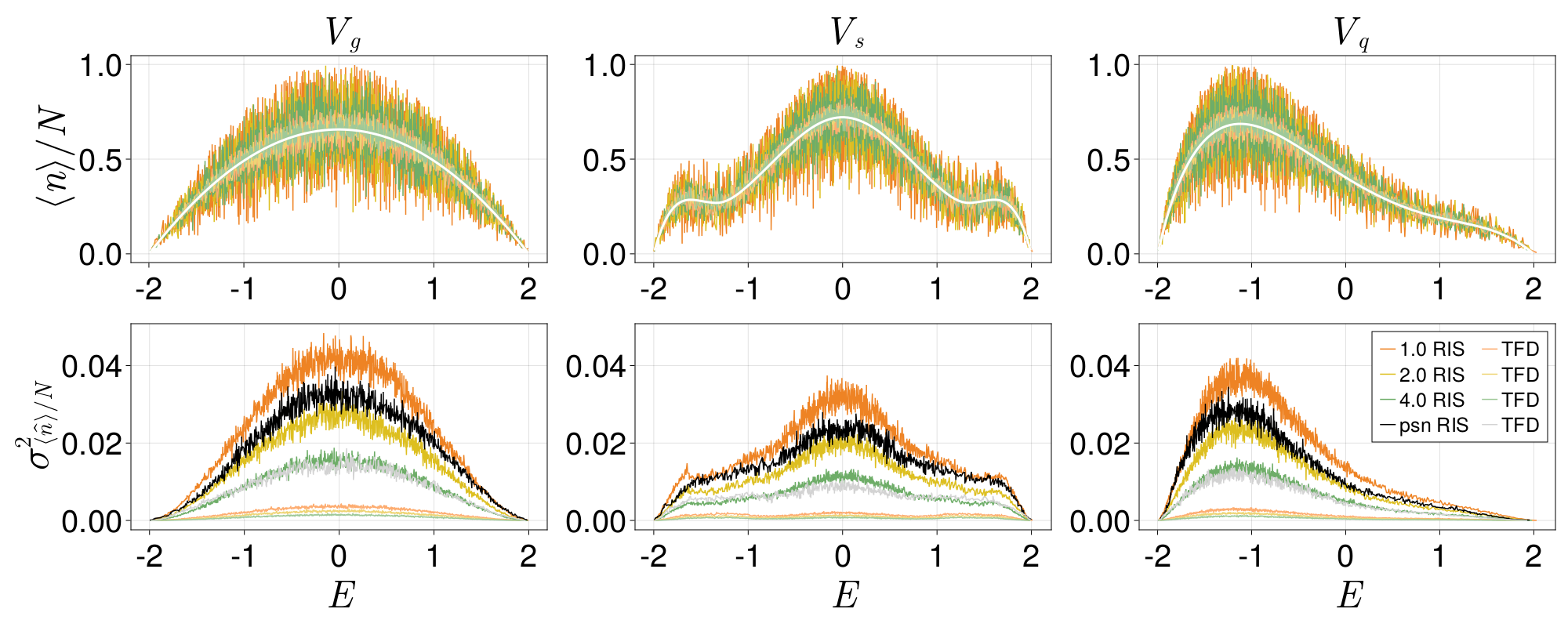}
    \caption{Top row:  Eigenstate complexity divided by system size relative to Random Initial States (darker colors, labeled RIS) and TFD initial states (lighter colors) for $N=1024$ random matrices with the potentials \eqref{pot} indicated in the title.  The potentials were normalized so that energy eigenvalues range from $-2$ to $2$.  We considered ensembles with various Dyson indices ($\beta_D=1, 2, 4$) or Poisson (psn) distributed  spectra (labeled by colors). The analytical average value obtained from \eqref{analytical_eig_comp} is the solid white line.  Bottom row: The variance of the eigenstate complexity, over $256$ draws of random matrices from the indicated ensemebles. }\label{figecom}
\end{figure}

In the top row of Fig. \ref{figecom} we compare this formula for $\overline{C_E}$ with the exact numerical computation of $C_E$ in RMT for the three potentials studied throughout the article, with initial TFD and random states for the definition of the Krylov basis.
We also plot the variance of the eigenstate complexity, here estimated as the variance of the complexity $C_{E_i}$ of the $i$th largest eigenstate, plotted against the mean value of $E_i$; here we use the ensemble variance rather than the local variance as it is cleaner to compute numerically and does not make a difference in the RMT case that we are studying here.  While the mean eigenstate complexity only depends on the density of states and so is the same whether we pick a random initial state or one with continuous support, the variance of eigenstate complexity does differ between the two types of initial states and different Dyson indices $\beta_D$.

We showed above that for an initial state with continuous $P(E)$, the plateau in spread complexity is completely determined by the mean Lanczos coefficients in the large $N$ limit. In particular it is not affected by the variance in these quantities and therefore by the variance in eigenstate complexity. But this is not the case for states with noisy $P(E)$ in the large $N$ limit, for example random initial states. The plateau of the spread complexity in general is given by 
\be    
    \bra{\psi(t)}\hat{n}\ket{\psi(t)}= \sum_{E,E'}\braket{\psi(t)}{E}\bra{E}\hat{n}\ket{E'}\braket{E'}{\psi(t)}\;,
\ee
and since the phase averaging trick \eqref{phase_avg} applies to all states regardless of whether $P(E)$ is continuous, after a long time average of the above, only the $E=E'$ terms contribute and so 
\be
    C_\text{plateau}=\sum_{E,E'}\braket{\psi}{E}\bra{E}\hat{n}\ket{E}\braket{E}{\psi}=\sum_E P(E) \, C_E\;.
\ee
If $P(E)$ is continuous, then over small intervals $[E,E+dE]$, $P(E)$ is approximately constant, $C_E$ can be replaced with its local mean $\overline{C_E}$, and so the plateau may be accurately analytically estimated by
\be
    C_\text{plateau}=\sum_E P(E)\, \overline{C_E} \approx \int \rho(E)dE\, P(E)\, \overline{C_E}\;.\label{smooth_state_plateau}
\ee
However, when $P(E)$ is noisy, there is the possibility that the noise in $P(E)$ is correlated with the noise in $C_E$, and so the plateau can differ from \eqref{smooth_state_plateau}. For random initial states in RMT, and even for systems with Poisson distributed eigenvalues, the noise in $P(E)$ must be negatively correlated with the noise in $C_E$, as we can see from the lowered plateau in Fig.~\ref{TFDSC2}, to be described later.

This negative correlation can be traced back to Anderson localization: the variation in the Lanczos spectrum causes the energy eigenstates to be localized in the 1-D chain near their average values, and so the further away the eigenstate's center is from the beginning of the chain, the smaller its overlap with the initial state (which is just the wavefunction at the beginning of the chain). Furthermore, the local variance of the eigenstate complexity 
is itself a good indicator for how localized the eigenstates are along the chain: \eqref{energy_slice_density} gives the sum of $\abs{\braket{E'}{K_n}}^2$ for the eigenvalues $E'$ in the interval $[E,E+dE]$.
Since the distribution over $n$ on the right hand side of \eqref{energy_slice_density} is the same regardless of the localization of the eigenstates in the Krylov basis, the more localized the eigenstates are in $x$, the more their individual means must vary in order for them to cover the same distribution, and hence the variance of the eigenstate complexity is \emph{larger}.
This is illustrated in the trend in the variance in the eigenstate complexity across Dyson indices in Fig.~\ref{figecom}; as $\beta_D$ decreases, the variance in the Lanczos coefficients increases \eqref{cov}, so the eigenstates are more localized, hence the variance of the eigenstate complexity increases. The role of Anderson localization in the Krylov chain has been discussed for other reasons in \cite{Rabinovici:2021qqt, Rabinovici:2022beu}.

\begin{figure}
    \centering
    \includegraphics[height=6cm]{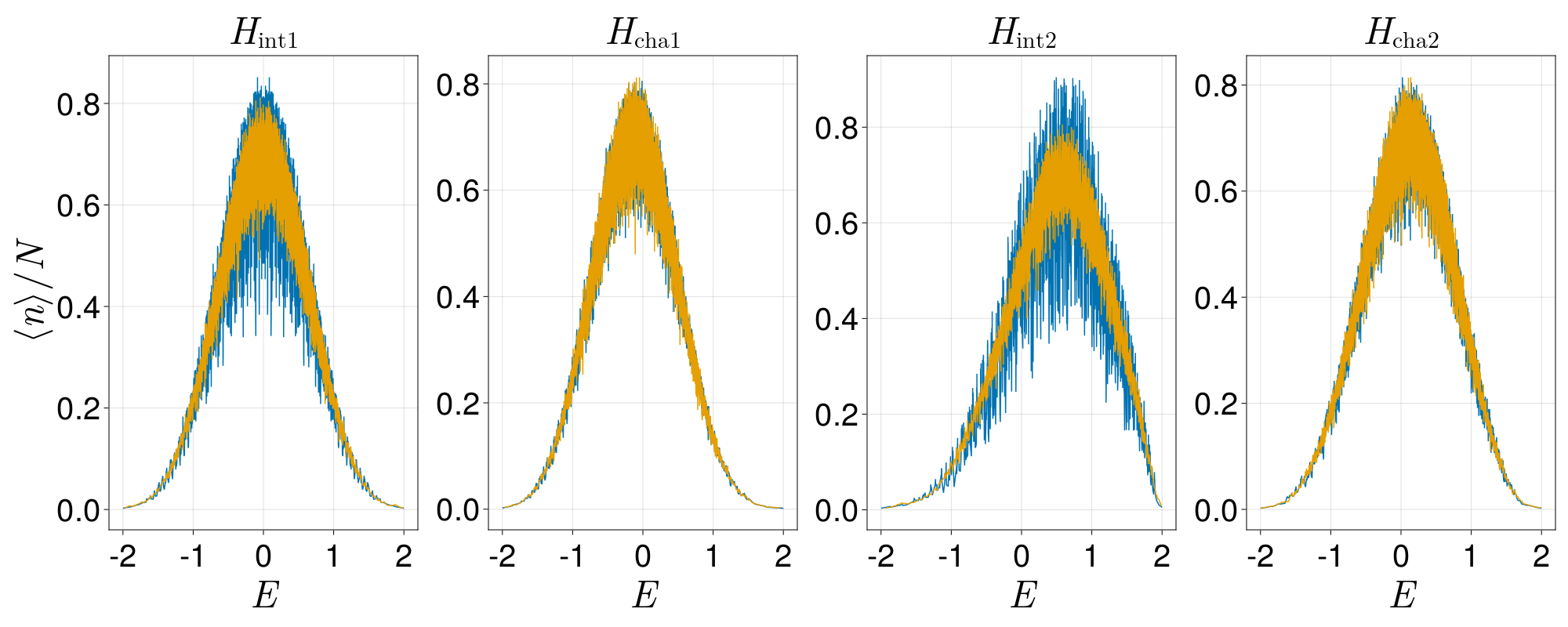}
    \caption{The scaled eigenstate complexities of the  Hamiltonians indicated in the panel titles (blue) from \eqref{spin_chain_ham}, for a TFD initial state, compared to the scaled eigenstate complexities of random matrix Hamiltonians with the same density of states (orange). We see that realistic chaotic systems conform with random matrix theory predictions of the variance of the eigenstate complexity while integrable systems do not.}\label{eigc_int_cha}
\end{figure}

Lastly, we show numerically that while the average of the eigenstate complexity only depends on the density of states, the variance is quite sensitive to the difference between integrability and chaos. In Fig.~\ref{eigc_int_cha} we plot the eigenstate complexity with an infinite temperature TFD initial state for the examples of chaotic and integrable systems in \eqref{spin_chain_ham}. As with the Lanczos spectrum itself, we see that while the variance in chaotic systems perfectly matches RMT, integrable systems result in larger variances.\footnote{We chose to show a TFD initial state here because the difference between chaotic and integrable systems is most visibly obvious there, as opposed to random initial states. This is consistent with comparisons of eigenstate complexity between RMT and Poisson systems in Fig.~\ref{figecom}.} 

\section{Universality classes and the shape of the spread complexity slope}\label{SecV}
In the context of quantum chaos, universality refers to  behavior which does not depend on the specific potential defining the RMT, and which is shared by more realistic chaotic systems as well. This behavior depends only on the symmetries defining the universality class. For example, many chaotic systems lie in the same universality classes as the Gaussian Orthogonal, Unitary, or Symplectic ensembles (GOE, GUE, and GSE). The universal behavior is encoded in several quantities, such as the level spacing and the covariance of the density of states, see \cite{Guhr:1997ve,akemann2011oxford} for detailed accounts. One dynamical quantity that carries the imprint of the universality class is the spectral form factor defined above. The spectral form factor is basically the Fourier transform of the two-point function of the density of states, so the fact that it codifies the universality classes is expected. Qualitatively, as described in detail in \cite{Cotler:2016fpe}, the spectral form displays four  regimes. There is a downward slope to a ``dip'', followed by an upward ramp to a plateau. This is qualitatively the same behavior we found for the survival probability of the initial random state, see Fig.~\ref{RISSA-bad}. In fact, the spectral form factor is a survival probability itself. It is the survival probability of the time evolved TFD state $SFF=\vert\langle\psi_{\beta+2it}\vert\psi_{\beta}\rangle\vert^2$. This observation has been used in different directions by several authors \cite{Papadodimas:2015xma,delCampo:2017bzr,Verlinde:2021jwu,Stanford:2022fdt,SpreadC}. For the spectral form factor, or for the survival probability associated with an initial state, the universality class of a RMT shows up in the nature of the transition from the ramp to the plateau. Indeed, since the leading density of states of the GOE, GUE and GSE classes is the same, the downward slope does not distinguish them, and one has to wait very long times to see the distinctions. But if we are ready to wait for such long times, then for the GOE universality class the transition from the ramp to the plateau is smooth, for the GUE it is sharp and for the GSE it shows a kink \cite{Guhr:1997ve}.

\begin{figure}
    \centering
    \includegraphics[height=6cm]{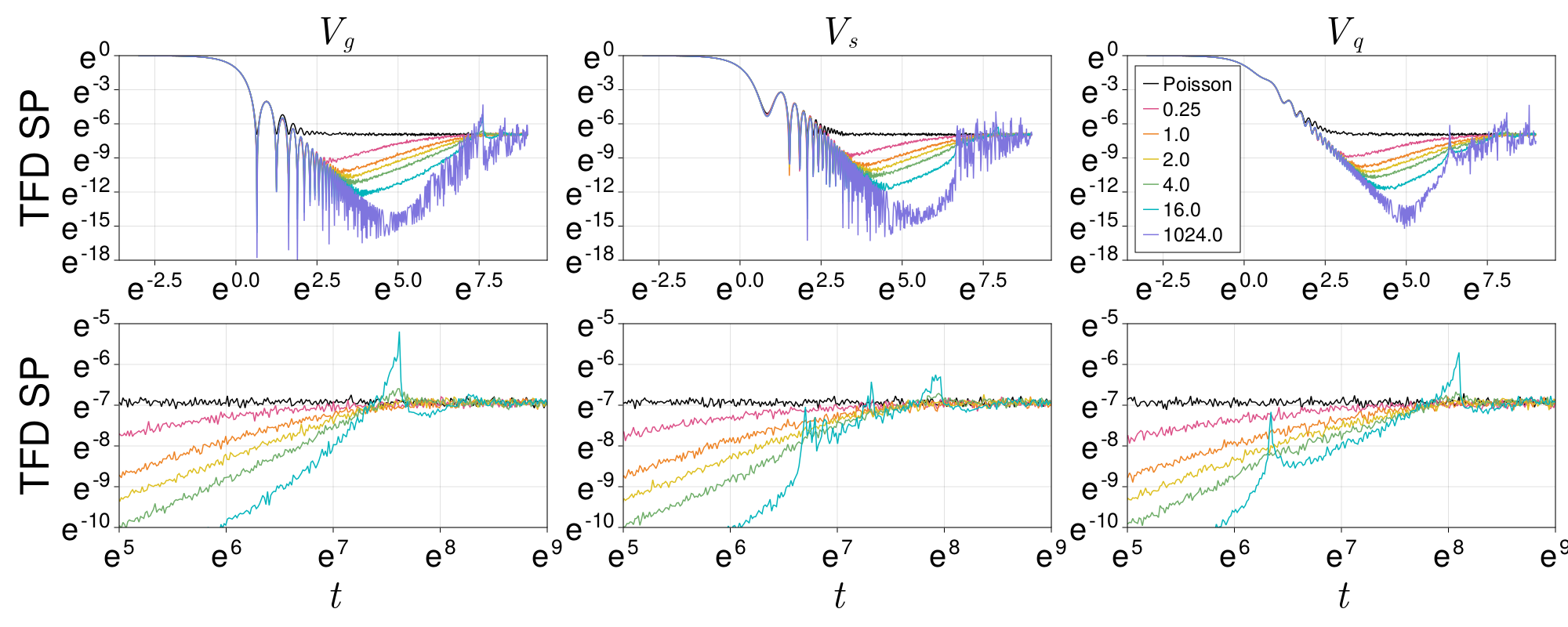}
    \caption{Survival probability of a maximally entangled state (i.e., spectral form factor) for $N=1024$ random matrices with the potentials indicated the panel titles, averaged over $256$ draws from the RMT ensemble. The Dyson index indicated by the color in the legend.  The black line is for a Hamiltonian with a Poisson distributed spectrum. Zoomed-in plots of the transition to the plateau are provided in the bottom row. We see that the dependence on the Dyson index lies in the transition from the ramp to the plateau.
    }\label{TFDSP2}
\end{figure}

\begin{figure}
    \centering
    \includegraphics[height=6cm]{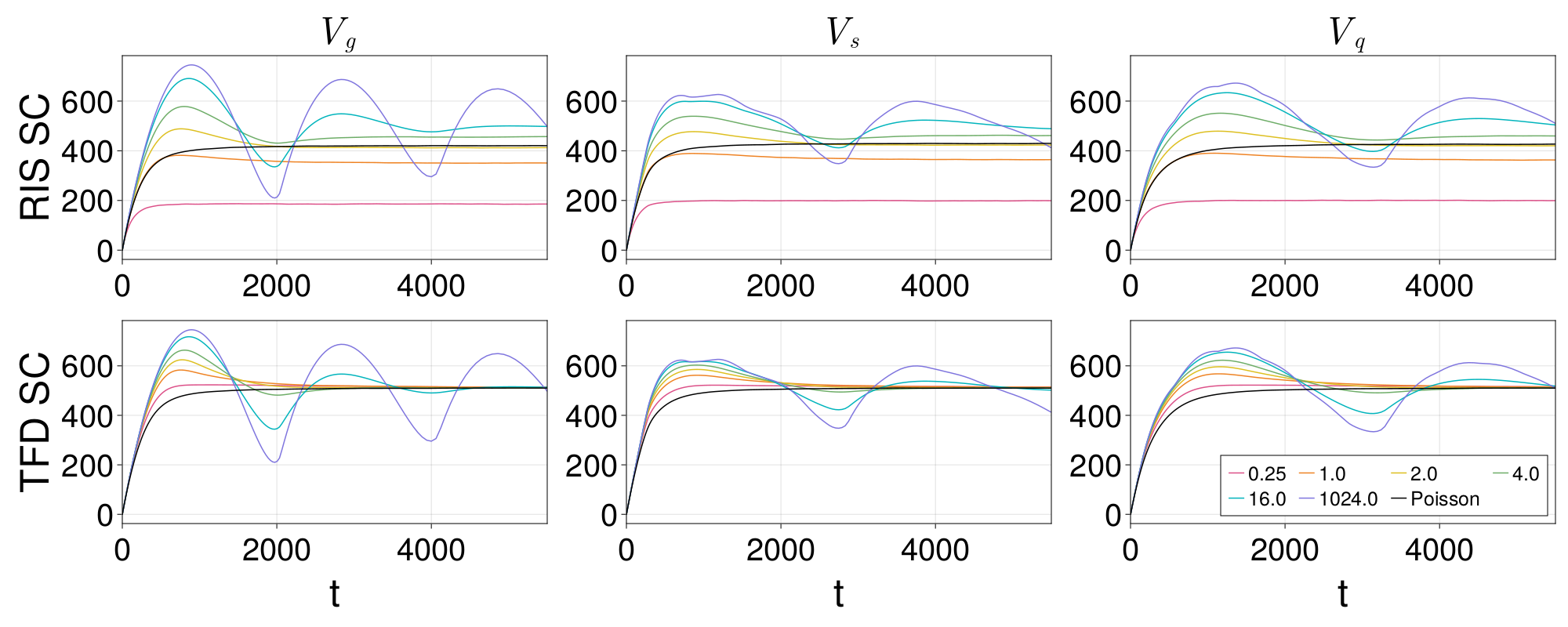}
    \caption{Spread complexity of random initial states (top row) and maximally entangled (TFD) states (bottom row) for $N=1024$ random matrices with the potentials indicated in the panel titles, averaged over $256$ draws from the RMT ensemble, and Dyson index indicated by the color in the legend. We see that the dependence on the Dyson index shows up in the damping of the oscillations of the spread complexity, which maintains the same shape between the random initial state and TFD cases.
    }\label{TFDSC2}
\end{figure}

It is interesting to ask if quantum complexity is sensitive to the universality class of the chaotic system. Since the spread complexity is a functional of the survival amplitude, one might hope that it will also codify the universality class, as was observed in \cite{SpreadC} for the GOE, GUE, and GSE. It turns out that in chaotic systems the spread complexity displays analogous regimes to the spectral form factor, namely a linear upward ramp to a peak, followed by a downward slope to a plateau (Fig.~\ref{RISSC}). The authors of \cite{SpreadC} noted that the downward slope was the key feature of the spread complexity distinguishing chaotic theories from ones with a Poisson spectrum.
The universality class of the RMT shows up in the nature of the transition between the slope and the plateau. This observation was further supported in \cite{Balasubramanian:2022dnj,Erdmenger:2023shk}. Here we expand on this connection by analyzing the transition between the slope and plateau for the generalized Gaussian beta ensembles introduced in Sec.~\ref{RMT_trid}. 
Using our analytical formulae for the mean and variance of the Lanczos coefficients \eqref{eq:one_point_final} and \eqref{cov}, along with the methods of Sec.~\ref{trid_predicts_SFF}, we can sample tridiagonal matrices whose spectra approximately match \eqref{jointb}
\be
p(\lambda_1,\cdots ,\lambda_n)=Z_{\beta_D,N}\, e^{-\frac{\beta_D\,N}{4}\sum_{k}V(\lambda_k)}\,\prod_{i<j}\vert \lambda_i-\lambda_j\vert^{\beta_D}\;,\nonumber
\ee
for any positive real $\beta_D$. Exact results for the distribution of the Lanczos coefficients for quadratic potentials $V(E)$ can be found in \cite{Dumitriu_2002}.\footnote{As an alternative approach we may sample these and stretch the spectra as described in Sec.~\ref{trid_predicts_SFF}; this method was not used in the present figures, but the results are the same.}
We then use these ensembles to compute the survival probability and spread complexity associated with the evolution of the TFD and the random initial state, all for different values of $\beta_D$. This is shown in Fig.~\ref{TFDSP2} and Fig.~\ref{TFDSC2} respectively.  The same features that appear in the survival amplitude and the spectral form factor in the transition from the ramp to the plateau show up as well in the transition from the complexity slope to the plateau. These numerical experiments lead us to conclude that the long time behavior of spread complexity codifies the universality class of chaotic systems.

But the spread complexity, and more precisely the one dimensional motion in the Krylov basis, gives a different physical picture of what is going on as we vary the universality class through the Dyson index $\beta_D$. In the Krylov picture, the survival probability is just the first entry of the probability distribution associated with the Krylov basis. This distribution has $N$ entries, where $N$ is the dimension of the Hilbert space explored by the time evolution. The slope of the spread complexity appears because for chaotic theories the one-dimensional hopping dynamics maintains the coherence of the probably distribution in the Krylov basis over long time scales. This means that by the time the mean position on the Krylov chain arrives for the first time at the plateau value, the probability distribution is still very narrow and stationarity has not yet been reached. The mean position then goes beyond the plateau until it bounces back. This behavior can repeat itself leading to oscillations in the spread complexity,  as we see for higher values of the Dyson index, in particular already for the GSE case $\beta_D=4$. This behavior was first described in \cite{SpreadC} and strong analytical evidence was given in \cite{Erdmenger:2023shk}.

We conclude that the oscillating nature of the transition between the spectral form factor ramp to the plateau, or the spread complexity slope to the plateau, is controlled by the amount of coherence in the one-dimensional hopping dynamics appearing in the Krylov description. In turn, this coherence is related to the size of the noise in the Lanczos spectrum around the analytical mean computed from RMT for a given density of states. Intuitively, we expect the lower the noise, the higher the coherence, and the larger the oscillations before the system settles onto the plateau. 
This expectation can be seen to hold directly from the dependence on $\beta_D$ of the covariance of the Lanczos coefficients \eqref{cov}. Interestingly, as seen in Fig.~\ref{TFDSC2}, the shape of the spread complexity remains the same relative to both random initial states and TFD states, even though the covariances of the Lanczos spectrum relative to these states are markedly different. It would be interesting to further study the relation between universality classes, coherence of the wavefunction in the Krylov basis, and the statistics of the Lanczos spectrum.

Summarizing, the complexity slope codifies the universal correlations of the chaotic Lanczos spectra for all initial states (Fig.~\ref{TFDSC2}). It plays the role of the ramp in the spectral form factor. That said, note that the complexity plateau is not by itself a clean smoking gun for chaotic dynamics. As we showed above, the plateau value depends on the initial state and, for a fixed density of states, could even be larger for Poisson spectra than for chaotic spectra in some cases.

\section{Discussion}\label{SecC}
In this article we have put forward a conjecture that quantum chaotic systems display a Lanczos spectrum whose statistics are well-described by Random Matrix Theory. Using new analytical and numerical tools, we have demonstrated, in some examples of chaotic systems, that both the variance of the Lanczos coefficients and the variance of the eigenstate complexity match those of random matrices.  We also developed a method for writing down random matrix theories with the same mean density of states as an integrable theory, and showed that the integrable theory will have much larger variances for its Lanczos coefficients.
Our conjecture complements the famous proposal of Bohigas-Giannoni-Schmit \cite{PhysRevLett.52.1} that the spectrum of chaotic Hamiltonians is well approximated by RMT.  The BGS conjecture is sometimes understood to mean simply that a chaotic theory will have the spectral density and eigenvalue repulsion implied by the Wigner surmise.  But the spectra of chaotic theories actually has additional correlations.  These correlations affect the Lanczos spectrum (see Fig.~\ref{varlz_poisson}) and hence the late time dynamics of theory.  We showed that by directly constructing the Lanczos spectra of RMTs we capture all the  necessary correlations to describe this late time behavior (see Figs.~\ref{RISSA-bad} and \ref{RISSC}).

We also showed that certain aspects of the Lanczos spectrum encode the long time dynamics, in particular the stationary distribution of the time evolution of the initial state in the Krylov basis. For initial states with smooth support in the energy basis, we showed that this plateau only depends on the density of states and cannot distinguish between integrable and chaotic systems.  Random initial states, which do not have smooth support in the energy basis, however, do generate plateaus that distinguish different scenarios.  This observation prompted us to study the complexity of eigenstates of the Hamiltonian relative to a given initial state, defined in terms of their mean support on the associated Krylov basis. As it turns out, the local variance of the complexity of these eigenstates strongly distinguishes integrable and chaotic systems, when associated with an initial state with continuous energy support. We connected this fact to Anderson localization in the Krylov chain. 

In all cases we have studied, the  Lanczos coefficients are ultimately functions of  the eigenvalues of the Hamiltonian. For example,  an initial state of the form $\sum_E \sqrt{P(E)}\ket{E}$ can be written in terms of the energy basis and can be evolved by the diagonal Hamiltonian to obtain the Lanczos coefficients. Likewise, for a random initial state, changing to the energy basis will not change the statistics of the state and again, we can simply evolve a random state by the diagonal Hamiltonian to obtain the Lanczos coefficients. The strength of the tridiagonal point of view is not that it contains more information than the spectral statistics, but rather that it transforms this information into a format that is easily analyzed, so that the difference between the statistics of chaotic and integrable systems is faithfully and directly represented in the covariance of the matrix entries. 
One potential avenue to explore further is how much more information  the covariance of the Lanczos coefficients can give us? We know that it can show a qualitative difference between different integrable systems, such as in Fig.~\ref{intcha_mean_variance}. If we do not mix the different sectors of the theory with the one-site disorder in our Hamiltonians in \eqref{spin_chain_ham}, the variance of the Lanczos coefficients will not match the random matrix results. Is it possible to pin down what aspects of the symmetries of those theories give rise to these differences?

While the analytical results in \eqref{eq:one_point_final} and \eqref{cov} have proven very useful, this type of analysis may be able to go further if we had analytical results for the covariance of the Lanczos spectrum relative to states other than random initial states, or Hamiltonians other than random matrices. Several interesting facts appear in our numerics, but as of yet remain unexplained: What exactly causes the difference in the variances in $a_n$ and $b_n$ for Poisson and Wigner-surmise-nearest-neighbor distributions of eigenvalues at the end of the Krylov chain in Fig.~\ref{varlz_poisson}? Why do random matrices differ so much in eigenstate complexity from Hamiltonians with Poisson distributed eigenvalues and integrable systems when we start with an initial state with continuous support over the energy basis (as shown in  Figs.~\ref{figecom} and \ref{eigc_int_cha}), but  not  nearly as much when we start with a random initial state? Why does the shape of the complexity curve in Fig.~\ref{TFDSC2} not depend on the choice of initial state? We have put the mean Lanczos coefficients, the mean eigenstate complexity, and the long time plateau probability over the Krylov chain under analytical control, but it would be useful to achieve more analytical traction for the covariances and for the approach towards long times in the dynamics.

It would also be interesting to explicitly apply our methods to the SYK model \cite{kitaev,sachdev,sachdev2} and its double scaling limit \cite{Berkooz:2018qkz,Berkooz:2018jqr,Lin:2022rbf,Berkooz:2022mfk} because these theories are 
more realistic as models of quantum black holes and quantum chaos, and nevertheless remain solvable to a large extent. Interestingly, the ``chord diagram'' approach  \cite{Berkooz:2018qkz,Berkooz:2018jqr} used to solve these theories is related to the Krylov methods discussed above \cite{Lin:2022rbf,Rabinovici:2023yex}. In this approach, with a suitably chosen initial state and Hamiltonian, the Krylov states $\ket{K_n}$ correspond to geometries in the dual 1+1-dimensional gravity that have  definite length. The studies demonstrating these points were essentially performed in the limit $n,t\ll N$ and so do not probe the descent of the Lanczos coefficients to zero at the tail of the Krylov chain, and therefore the long time behavior discussed in the present paper.  From the perspective of the SYK model and its doubling scaling limit, the descent of the Lanczos coefficients is believed to be controlled by non-perturbative effects,  described in the dual quantum gravity theory by higher genus contributions to the path integral, i.e., wormholes. Our methods may be useful for analyzing such non-perturbative effects, where there is an interplay between  long time dynamics, spreading of the wavefunction on the Krylov chain, and wormholes in the quantum gravity path integral.

Our methods may also shed light on the information loss paradox for black holes. One formulation of the puzzle observes that the unitarity of quantum mechanics in a black hole background is challenged by the apparent decay to zero of correlation functions of quantum fields in a black hole background at large time separations \cite{Maldacena:2001kr}.  On general grounds, this decay is not possible via unitary dynamics in a finite entropy system.  The analysis described in this paper gives an analytic understanding of why, in chaotic theories like those expected for quantum gravity, such correlations will initially decay but will stabilize to a non-zero long-term average with fluctuations of calculable magnitude.  Mechanistically,  the approach to this dynamical steady state is controlled by the descent of the Lanczos coefficients to zero at the tail of the Krylov chain.  To study this in detail for a realistic quantum gravity, we could examine a conformal field theory dual to gravity in AdS space.  To apply our methods to such a continuum theory we would need to appropriately restrict the analysis to the part of the spectrum that is actually explored in the dynamics of a process, e.g., by working in the microcanonical ensemble.  Alternatively we could consider the Matrix model of M-theory in either the original kinematic interpretation \cite{Banks:1996vh} or the alternative understanding in terms of a duality \cite{Balasubramanian:1997kd, Polchinski:1999br}.  In this case, because we are working with matrix dynamics, the methods of the present paper may be more readily applicable, now to the gravitational theory in 11 dimensions from which all the string theories are supposed to descend.

More generally, to completely solve the black hole information paradox we need to understand how initial states like elephants and encyclopedias collapse to make black hole microstates which are nearly impossible to probe through standard means, thus giving the impression of being hidden in a causally disconnected region, and then evaporate into radiation that remembers the initial collapsing state. A simple explanation for why the microstates are so difficult to probe is that they are very complex so that standard observables like low-order correlation functions of simple operators give nearly universal responses \cite{Balasubramanian:2005kk} that require great precision, or long times \cite{Balasubramanian:2006iw} to resolve.  The behavior of the spread complexity in chaotic systems may quantitatively explain why and how this happens: we showed that there is a precise sense in which unitary chaotic dynamics  starting from simple initial states with smooth support in the energy basis reaches a maximally complex dynamical steady state that is evenly dispersed at late times over the Krylov basis.  We expect that low-order correlation functions in such late time states will be nearly universal,  with tiny time-dependent deviations that are sensitive to dynamical fluctuations in the state.  On the other hand, it is possible that there are very complex observables that can identify the late time state despite its high spread complexity, albeit at huge computational cost.  It would be interesting to identify such observables.

\medskip
{\bf Acknowledgements.} We are grateful to Pawel Caputa for many discussions about spread complexity and the Lanczos approach, and Anatoly Dymarsky, Martin Sasieta and Julian Sonner for conversations about quantum chaos. The work of VB and QW is supported by a DOE through DE-SC0013528,  QuantISED grant DE-SC0020360, and the Simons Foundation It From Qubit collaboration (385592). The work of JM is supported by CONICET, Argentina. VB thanks the Isaac Newton Institute for Mathematical Sciences for hospitality during the program ``Bridges between holographic quantum information and quantum gravity'', supported by EPSRC Grant EP/R014604/1.
\newpage
\bibliographystyle{utphys}
\bibliography{main}

\end{document}